# Theory of High-$T_C$ Superconductivity: Transition Temperature

**Dale R. Harshman**[1,2,3,6], **Anthony T. Fiory**[4] **and John D. Dow**[3,5]

[1]*Physikon Research Corporation, Lynden, WA 98264, USA;*
[2]*Department of Physics, University of Notre Dame, Notre Dame, IN 46556, USA;*
[3]*Department of Physics, Arizona State University, Tempe, AZ 85287, USA;*
[4]*Department of Physics, New Jersey Inst. of Technology, Newark, NJ 07102, USA;*
[5]*Institute for Postdoctoral Studies, Scottsdale, AZ 85253, USA*

**Abstract**

It is demonstrated that the transition temperature ($T_C$) of high-$T_C$ superconductors is determined by their layered crystal structure, bond lengths, valency properties of the ions, and Coulomb coupling between electronic bands in adjacent, spatially separated layers. Analysis of 31 high-$T_C$ materials (cuprates, ruthenates, rutheno-cuprates, iron pnictides, organics) yields the universal relationship for optimal compounds, $k_B T_{C0} = \beta/\ell\zeta$, where $\ell$ is related to the mean spacing between interacting charges in the layers, $\zeta$ is the distance between interacting electronic layers, $\beta$ is a universal constant and $T_{C0}$ is the optimal transition temperature (determined to within an uncertainty of $\pm 1.4$ K by this relationship). Non-optimum compounds, in which sample degradation is evident, e.g. by broadened superconducting transitions and diminished Meissner fractions, typically exhibit reduced $T_C < T_{C0}$. It is shown that $T_{C0}$ may be obtained from an average of Coulomb interaction forces between the two layers.

---

## 1. Introduction

High transition temperature (high-$T_C$) superconductors [1] are characterized by a layered, two-dimensional (2D) superconducting condensate and unique features [2] that are very different from conventional superconducting metals [3]. The two-dimensionality is evidenced by large anisotropies in the normal and superconducting electronic transport, and non-metallic transport in the direction perpendicular to the layered crystal structure. Superconducting properties can be optimized via doping or applied pressure to yield highest transition temperature and bulk Meissner fraction. Other materials properties that prove to be important from technological as well as scientific perspectives are relatively poor malleability and tendencies towards fracture, because ionic forces dominate crystal bonding [4] (and references therein). Since their discovery [1], the classification of high-$T_C$ superconductors has broadened to include the family of cuprates such as $YBa_2Cu_3O_{7-\delta}$, rutheno-cuprates (e.g. $RuSr_2GdCu_2O_8$), ruthenates such as $Ba_2YRu_{1-x}Cu_xO_6$ and $Sr_2YRu_{1-x}Cu_xO_6$, certain organic compounds (e.g. κ–[BEDT-

[6] Corresponding author: drh@physikon.net

TTF]$_2$Cu[NCS]$_2$ and κ–[BEDT-TTF]$_2$Cu[N(CN)$_2$]Br), various iron pnictide (and related) superconductors [e.g. La($O_{1-x}F_x$)FeAs], and possibly transuranics (e.g. $PuCoGa_5$). Optimal transition temperatures span a range from ~10 K to ~150 K.

Suppositions of pairing mechanisms based upon lattice vibrations in the high-$T_C$ superconductors have led to serious contradictions with experiment [5-11], and thus will not be considered here. Instead we focus on the systematic correlation between $T_C$ and 2D carrier concentration $n_{2D}$, unique to optimally doped high-$T_C$ compounds [2], indicating an electronic pairing mechanism. Proving that pairing in the high-$T_C$ compounds is Coulombic (electronic) in nature requires one to show that there exists a dependence of $T_C$ on specific charge and structural parameters that can be related to an electronic energy scale. To uncover such an intrinsic trend in $T_C$, it is important to consider only the optimum compounds, as was demonstrated previously [2]. Non-optimal compounds are typically identified by a clear degradation of the superconducting phase exhibited by a transition temperature depressed below the optimal $T_C$, hereinafter denoted as $T_{C0}$. Indicators that substituting cation impurities into optimum compounds may produce an impure superconducting state are, e.g. a broadened superconducting transition width $\Delta T_C$, an incomplete or suppressed Meissner fraction, and/or an enhanced oxygen isotope effect (OIE) [5,6,12].

A particularly relevant structural trait common to high-$T_C$ superconductors is the presence of at least two different types of charge layers. This is found in the cuprates, where the superconducting state is created by substituting ions of different valences in different layers. A well known example of cation structure doping is $Sr^{+2}$ substitution for $La^{+3}$ in $La_{2-x}Sr_xCuO_{4-\delta}$; an example of anion structure doping is $Ca^{+2}$ substitution for $Y^{+3}$ in $(Y_{1-x}Ca_x)Ba_2Cu_3O_{7-\delta}$; and an example of both is the compound $(Pb_{0.5}Cu_{0.5})Sr_2(Y_{0.6}Ca_{0.4})Cu_2O_{7-\delta}$.

Numerous manifestations of this electronic duality are displayed in several measurements. Hall and Seebeck coefficients (in the normal state) are used to distinguish whether cuprate or pnictide superconductors are n- or p-type (e.g. n-type $LaO_{0.89}F_{0.11}FeAs$ [13] and p-type $Ba_{0.6}K_{0.4}Fe_2As_2$ [14]); significantly, hole and electron carriers have been determined to coexist in cuprates of both types (n-type $Nd_{1.85}Ce_{0.15}CuO_{4\pm y}$ [15], see also [16,17][7]; p-type $YBa_2Cu_4O_8$ [18], $YBa_2Cu_3O_{7-\delta}$ and $Bi_2Sr_2CaCu_2O_{8+\delta}$ [19]). Fermi surfaces contain sheets with both electron-like and hole-like character in the doped FeAs-based superconductors [20] and in the transuranic superconductor $PuCoGa_5$ [21]. In the cuprates there are bands of carriers associated with $CuO_2$ planes as well as a second layer type (e.g. BaO-CuO-BaO in $YBa_2Cu_3O_{7-\delta}$ from thermoelectric power [22]; Bi-O(2)-O(3) in $Bi_2Sr_2CaCu_2O_{8+\delta}$ from band structure and confirmed by O-1s absorption-edge spectroscopy [23]). In the case of $YBa_2Cu_3O_{7-\delta}$ n-type behaviour occurs at $\delta = 0.79$ by substituting $La^{+3}$ for 13% of the $Ba^{+2}$ [24], while coexisting electron and hole Fermi sheets are observed for $\delta \approx 0.50$ [25]. Two electronic components are also evident in low-temperature thermal measurements in superconducting states [26-29]. [8]

These dualities in properties credibly demonstrate that the superconductivity involves the presence of two electronic bands, at the least, and that they are associated with different layer structures. Strengths of electronic coupling between layers in high-$T_C$ materials, as indicated by effective mass anisotropy $\Gamma = m_c^*/m_{ab}^*$ (for crystallographic axes *a*, *b*, and *c*), can differ by orders of magnitude between compounds with similar transition temperatures; e.g. for $YBa_2Cu_3O_{7-\delta}$ $\Gamma = 26$ [30], whereas for $Bi_2Sr_2CaCu_2O_{8+\delta}$ $\Gamma > 1300 - 3000$ is indicated [31,32], yet both compounds have $T_{C0} \sim 90$ K. Thus, charge-transfer coupling between layers appears relatively unimportant in determining $T_C$, although the magnitude of $\Gamma$ governs transport of electrons along the *c*-axis (hard-axis) and may influence charge equilibration.

[7] P-type doping via interstitial oxygen was suggested in [16], consistent with observed Ce-dependence [17].

[8] Excess specific heat ($\gamma_0 T$ term in p- and n-type, irrespective of pairing symmetry indications [26,27]) and offsets in thermal conductivity, $(\kappa/T)_{T\to 0} > 0$, observed in optimal high-$T_C$ superconductors at low-T, vary too strongly among compounds to be considered consistently as universal evidence of a noded gap [28,29].

Additionally, measured pairing state symmetries of high-$T_C$ superconductors can be either nodeless or not (e.g. indicators of both are found in the cuprates [27,33-39]),[9] which imparts a subsidiary role for wave-vector dependence of superconducting energy gaps in determining $T_C$. Other broad differences among the high-$T_C$ families are found in band structures and Fermi surface topologies (e.g. incomplete Fermi arcs are observed in the optimally doped cuprate superconductors [40]), which suggests that compound-dependent energy band and quasiparticle properties are also inessential. We deduce from these considerations that to zeroth order, the transition temperature of high-$T_C$ superconductors is found by properly identifying and quantifying the layers of charges. This leads to the theoretical approach developed in this work, which finds that the Coulomb interaction between charges in spatially separated layers yields high-$T_C$ superconductivity. Even for charges of opposite signs (i.e. electrons and holes in separate layers [19]), the ground state is not an excitonic insulator or dipolar superfluid [41] because symmetry is broken by disparities in scattering rates (which can be magnetic in origin), effective masses, and two dimensional confinement (e.g. in cuprates carriers of one sign type are dominant for the transport in the normal state [19], the penetration depth in the superconducting state [42], and the cross-over from two- to three-dimensionality with doping [43]).

Several published theoretical treatments have considered the formation of a superconducting state via interacting electron/electron (or electron/hole) bilayer systems. Of particular relevance is the work of Little [44] in which electrons in the $d_x2$ orbitals of the Cu atoms in the cuprate planes interact with electrons in the oxygen $p_z$ orbitals of the chain-oxygens (i.e. the apical, or BaO-layer oxygens of $YBa_2Cu_3O_{7-\delta}$). However, since there is no correlation of $T_C$ with the distance separating the apical oxygens and the planar Cu ions, this model misidentifies the interacting-ion orbitals. Nearly two decades earlier, Pashitskiĭ considered collective excitations between n- and p-type semiconductor layers (with identical work functions) in lamellar structures [45]. Assuming unequal effective masses $m_n << m_p$, Pashitskiĭ showed that for thin n-type (metallic) layers sandwiched between heavily-doped p-type layers, such an interaction can lead to the pairing of the electrons and 2D superconductivity via the exchange of virtual surface plasmons ($T_C$ estimated in the range $10^2$-$10^3$ K). These studies, though important, do not quite have the necessary attributes required for high-$T_C$ superconductivity. Other theoretical work published in 1976 [46,47] concerned "superconductivity" in systems with spatially separated electrons and holes; later it was clarified that these theoretical models were not superconducting, but treated neutral superfluidity in symmetric systems with counter flowing electrical currents [41].

In the example of $YBa_2Cu_3O_{7-\delta}$ the doping and temperature dependence of the thermoelectric power reveals association of oppositely signed components with $CuO_2$-Y-$CuO_2$ and BaO-CuO-BaO structures, i.e. spatially separated electron-like and hole-like carriers (similarly for $Bi_2Sr_2CaCu_2O_{8+\delta}$) [22]. Noting that holes are the dominant carriers of superconductivity [42], one deduces that electrons in separate layers serve to mediate the pairing, as discussed in [19]. To prove that this picture is generally correct for p-type and n-type (where the roles of the two carrier types are reversed) high-$T_C$ families, one must determine both the density of the superconducting carriers $n_{2D}$, and the separation distance between the superconducting and mediating layers. Since cation substitutions and other doping variations tend to degrade the superconductivity of erstwhile optimal compounds, we presume that highest $T_C$ ($\equiv T_{C0}$) occurs for a given charge-structure equilibrium.

In section 2 it is shown that logical rules can be applied to deduce carrier densities from the geometry of the lattice structure and valency properties of the ions. Where available, the results for $n_{2D}$ are shown to be validated independently by experiment. The analysis of the experimental data presented in this work shows that $T_{C0}$ is proportional to the square-root of the *2D superconducting interaction*

[9] Different techniques appear to be sensitive to different components of the superconducting state [27,33-39], and muon spin rotation ($\mu^+$SR) [33-36], in particular, can be affected by the pinning structure.

*density*, $1/\ell$, divided by the separation between adjacent electron and hole layers, denoted as the distance $\zeta$. The result is $k_B T_{C0} = \beta/\ell\zeta$, where $k_B$ is Boltzmann's constant and $\beta = 0.1075 \pm 0.0003$ eV Å$^2$ is universally constant among all the high-$T_C$ compounds. Explicit dependence of $T_{C0}$ on parameters related to band structure, such as bandwidths, effective masses, Fermi surface structure, and dielectric constants, appears to be either absent or subsumed within the result. Fundamentally, therefore, the transition temperature of high-$T_C$ superconductors is found to obey a remarkably simple algebraic dependence on the two length parameters that express the density of carriers in electron and hole layers and the separation between them.

The model of high-$T_C$ structure, the determination of carrier densities, and demonstration of the universality of the results are presented in section 2. The form of the interlayer Coulomb interaction is discussed in section 3. In section 4 we present a discussion regarding our model and the nature of the mediating band above and below $T_{C0}$, and our conclusion is presented in section 5.

## 2. Coulombic model of high-$T_C$ superconductivity and experimental test

For the purpose of developing our theory, we present a generalized model of high-$T_C$ superconductivity that is independent of locations of the two carrier types, as well as their roles in pairing and mediation. We identify type I reservoirs (associated with $e_I$ carriers) with the BaO-CuO-BaO (or equivalent) layers, specifying the number of *outer* (or interacting) layers by $\nu$ (i.e. $\nu = 1$ for single-layer reservoirs, $\nu = 2$ for multiple-layer reservoirs). Similarly, type II reservoirs (associated with $e_{II}$ carriers) are identified with the cuprate-plane (or equivalent) layers, with the *total* number of $CuO_2$ planes denoted by $\eta$ ($\eta = 1, 2, \ldots$). A representative model structure is depicted in figure 1. The elemental superconducting structure repeats with spatial periodicity $d$ as indicated. The interacting layers of the type I and type II reservoirs are separated by the spacing $\zeta$. Doping layers present in certain high-$T_C$ compounds are shown in figure 1 as sandwiched (diagonal hatching) between pairs of type I (depicted for $\nu = 2$) and type II (for $\eta \geq 2$) layers. Among the various high-$T_C$ families the number of interacting layers or the presence of separate doping layers can vary.

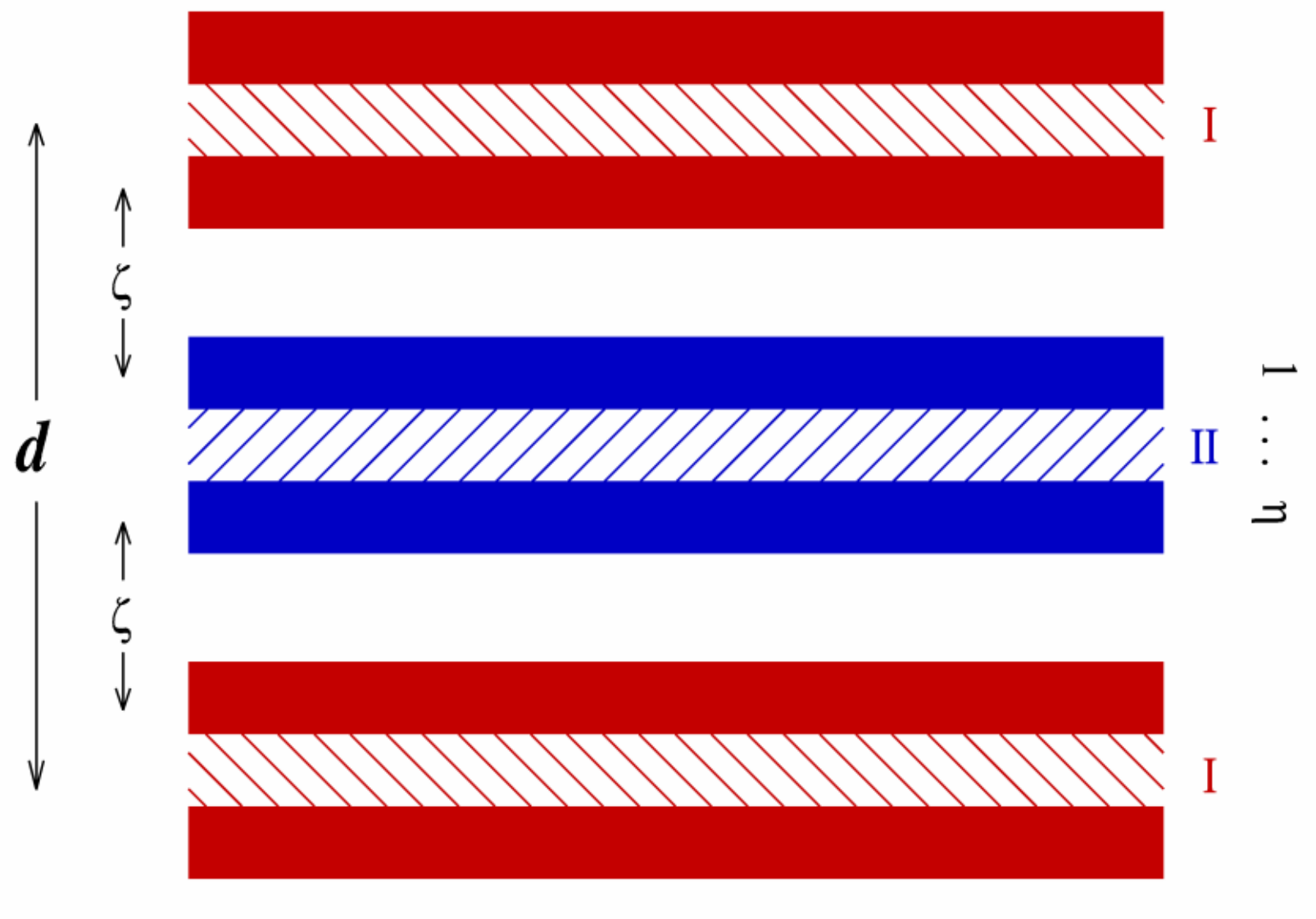

**Figure 1**. Representative model structure of high-$T_C$ superconductors. Cross section view perpendicular to basal plane of periodic electronic layers of types I (red, depicted here with $\nu = 2$) and II (blue). $\zeta$ is the separation between adjacent layers of opposite type; $\eta$ is number of type II layers; $d$ is the periodicity.

Hypothetically, the simplest high-$T_C$ superconducting structure would contain alternating single electron layers and single hole layers, where $\nu = \eta = 1$ [48]. Of the materials discussed herein, the "infinite-layer" compound $(Sr_{0.9}La_{0.1})CuO_2$ probably best illustrates this category. Most high-$T_C$ compounds have two outer type I layers ($\nu = 2$, often with intervening doping layers) comprising the type I charge reservoirs, with a 2D charge density per outer layer of $\sigma_I/A$ (where $\sigma_I$ is the fractional charge per (outer) type I layer and $A$ is the area of the basal-plane, both per formula unit). Thus $\nu$ defines the number of interacting layers in the type I reservoirs. For $\eta \geq 2$, the … $CuO_2$–(RE)–$CuO_2$ … structures form an extended $e_{II}$ reservoir, with a 2D charge density per layer of $\sigma_{II}/A$ ($\sigma_{II}$ is the fractional charge per type II layer per formula unit), where charge can transfer to the end planes as needed. One would thus expect the inner layer(s) (for $\eta \geq 3$) to be relatively depleted of carriers, leaving them insulating, or near-insulating, as has been observed [49]. In $YBa_2Cu_3O_{7-\delta}$, for example, the type I reservoir comprises the BaO–CuO–BaO structure (two BaO layers separated by a CuO doping layer), where the CuO chains can transfer charge to the two BaO end planes. Since one normally dopes the type I layers, allowing charge to also transfer to the type II planes, we define the superconducting interaction density as $\eta\sigma_I/A$, and $\ell = (\eta\sigma_I/A)^{-1/2}$ as the mean planar distance between the superconducting interaction carriers.

Within this context, $\zeta$ is taken to be the perpendicular distance (i.e. along the hard-axis) between nearest-neighbor mediating and superconducting layers. In the cuprates, for example, $\zeta$ is taken to be the *c*-axis distance between the Ba, Sr or equivalent ions and the nearest $CuO_2$-plane oxygen ion(s). In order to minimize any dependency on planarity of the layer ions, we systematically identify (where possible) the positive ions in the type I layers and the negative ions in the type II layers to define $\zeta$. We note that knowledge of specific band structure is unneeded in determining $\zeta$.

### *2.1. Electron and hole densities: equilibrium and limiting values*

Band structure calculations have shown that there exist bands crossing the Fermi level associated with both the $CuO_2$ planes and BaO-CuO-BaO (or equivalent) structures [20,50,34[10]]. Any carriers introduced by doping would necessarily fill both bands, and charge transfer (which is prevalent in the high-$T_C$ systems) would affect the balance of charge between the holes and electrons. Evidence of charge transfer along the *c*-axis can be found in positron annihilation experiments [51], in which the transfer of electrons away from the CuO chain regions in underdoped (twinned) single-crystals of $YBa_2Cu_3O_{7-\delta}$ is observed as the superconducting condensate forms below $T_C$; we interpret this behaviour as a tendency for the superconducting state to accommodate a balance between holes and electrons.[11] The optimal transition temperature, $T_{C0}$, for any given high-$T_C$ material structure is thus defined as the highest $T_C$ attainable, which presumably occurs at this charge equilibrium. This establishes a governing principle unique to this work; single-ion doping of a high-$T_C$ compound populates both hole and electron reservoirs; the charge density in the type I reservoirs $n_{2D}(e_I)$ balances that in the type II reservoirs $n_{2D}(e_{II})$, such that $n_{2D}(e_I)$ / $n_{2D}(e_{II}) = 1$ at $T_{C0}$.

These correlated systems are also dilute in the sense that $\ell$ for optimal materials is generally greater than $\zeta$. This is a necessary condition for superconductivity based on Coulomb interactions between two spatially separated charge layers; the attractive potential between the interacting charge planes must be greater than the in-plane repulsive interaction such that,

$$\zeta \leq \ell \,, \tag{2.1}$$

---

[10] W. Y. Ching and W.-Y. Rulis, band structure of $YBa_2Cu_3O_{7-\delta}$ (unpublished).

[11] In single-crystal κ–$[BEDT\text{-}TTF]_2Cu[NCS]_2$ (generally considered slightly overdoped [52]), where the positron density distribution favours the $Cu[NCS]_2$ anion layers, lifetime is seen to decrease roughly linearly below $T_C$ [53], indicating an increasing electron density in these layers as the sample is cooled.

which is also a mathematical representation of Little's criteria (a) and (b) in [44] for a strong excitonic (electronic excitation) interaction in a layered superconductor system. We propose that the maximum possible transition temperature occurs for $\zeta = \ell$, placing an upper-limit on $T_{C0}$. Altering the equilibrium between the electrons and holes results in a decrease in $T_C$ below $T_{C0}$, which is manifested as the peak observed in $T_C$ versus doping (and pressure) curves. The logic behind equation (2.1) is similar to that of equation (9) of [2], except that we now know that optimization can occur for $\zeta < \ell$, and $\zeta$ now properly defines the distance between the mediating and superconducting layers.

Considering that high-$T_C$ superconductivity would necessarily arise from Coulombic interactions between the electrons and holes across the distance $\zeta$, the transition temperature would follow the mediating energy scale (i.e. Fermi energy, plasmon energy, interlayer Coulomb potential, etc.) such that,

$$T_{C0} \propto \ell^{-p}\zeta^{-q} , \qquad (2.2)$$

where p and q are positive exponents. Our task is thus to look for validation of equations (2.1) and (2.2) in the experimental data; obtaining accurate values for $\zeta$ and $\ell$ is essential in order to construct and confirm a theoretical model for the high-$T_C$ pairing mechanism.

### *2.2. Determining ζ and ℓ*

The interaction distance $\zeta$ is generally defined to be the distance along the hard-axis separating the nearest-neighbor mediating and superconducting layers. In the cuprate compounds, the outer (interacting) type I layers contain cations such as $Ba^{+2}$, $Sr^{+2}$, $La^{+3}$; the type II (anion) layers contain the $CuO_2$ planes; and $\zeta$ is typically measured between the outer cations in the type I layers and the cuprate-plane oxygens. For the Fe-based pnictides, $\zeta$ is the hard-axis distance between the Ln (or Ba/K) and the As (or equivalent) ions. For the [BEDT-TTF]-based organic compounds, $\zeta$ spans the gap between the anion layers and the nearest-neighbor S ions of the [BEDT-TTF] molecule. Values for $\zeta$ are generally well defined from x-ray or neutron diffraction measurements and Rietveld refinement analysis.

Determining $\ell$ means determining $\sigma_I$ and *A*. Like for $\zeta$, accurate values for *A* are readily available; however, equally accurate *experimental* values for $\sigma_I$ for most of the high-$T_C$ compounds are more difficult to find. Assuming the optimal relationship $n_{2D}(e_I) / n_{2D}(e_{II}) = 1$ holds true, one can write for optimal compounds,

$$n_{2D} = \nu\, \sigma_I /A = \eta\, \sigma_{II} /A , \qquad (2.3)$$

where we note that for $\nu = \eta$, one has $\sigma_I = \sigma_{II}$. For convenience, we drop the subscript and use $\sigma$ to represent the fractional charge $\sigma_I$ for the remainder of this work.

Estimations of $\sigma$ for the "90 K" phase of $YBa_2Cu_3O_{7-\delta}$ (where $\sigma_{II} = \sigma_I$) lie in the range 0.21–0.25 [54]. From the perspective of the model structure put forth herein, we define the average number of carriers per layer for highest-$T_C$ stoichiometry as $\sigma$. For $YBa_2Cu_3O_{6.92}$ [55][12] ($T_{C0}$ = 93.7 K [56]) $\sigma$ can be determined by considering the oxygen x content above the minimum, $x_0$ = 6.35 [57], required for superconductivity. The total oxygen content associated with the superconductivity is then (6.92 – 6.35) = 0.57. Given a valence of –2 per oxygen ion, the total number of electrons available to dope the superconducting structure is 2 × 0.57 = 1.14. In $YBa_2Cu_3O_{6.92}$, there are five conducting layers containing oxygen; two $CuO_2$ layers, two BaO layers and one CuO layer (three associated with the type I structures and two associated with the type II). Invoking our principle that doping populates both layer types, the charge allocated per layer for $YBa_2Cu_3O_{6.92}$ is thus 1.14/5 = 1.14/(2×2.5) = 0.228 (where 2.5 is the average number of conducting layers per reservoir), which we denote as $\sigma_0$, thus identifying this material as a reference or standard with which other high-$T_C$ compounds can be compared. By applying the same argument as above to the oxygen-ordered "60 K" phase, $YBa_2Cu_3O_{6.60}$ [55][13] ($T_{C0}$ = 63 K [58]) the average charge allocated per layer is found to be $\sigma = 2(6.60 - 6.35)/5 = 0.1$. Alternatively, $\sigma$ for $YBa_2Cu_3O_{6.60}$, can be calculated by taking the ratio of the "active" oxygen content (i.e. the oxygens associated with superconductivity) of the 60 K to the 90 K phases such that $\sigma = [(6.60 - 6.35)/(6.92 - 6.35)]\ \sigma_0 \approx 0.439\ \sigma_0 = 0.1$. The two methods of calculating $\sigma$ for $YBa_2Cu_3O_{6.60}$ agree as they must, showing that it is possible to determine $\sigma$ for a variety of optimal high-$T_C$ materials by scaling from $\sigma_0$ (= 0.228).

[12] Structure: table II of [55], x = 6.93; $\zeta$ measured between $Ba_z$ and average of $O(2)_z$ and $O(3)_z$.

Below we detail two methods of determining the value of $\sigma$ for a variety of compounds; absolute values can be determined via charge allocation and valency scaling properties of a given compound, whereas relative values can be extracted by comparing such properties to $YBa_2Cu_3O_{6.92}$, which is taken as the standard. The proposed model structure provides a method of relating $\sigma$ for optimal materials to the crystallographic structure.

Of particular importance is that for several of the optimized high-$T_C$ cuprates, $\sigma = \sigma_0$, which necessarily implies that $\sigma_0$ represents an important characteristic value for these electrovalent ionic structures. This is true by definition for $YBa_2Cu_3O_{6.92}$, and by analogy for $LaBa_2Cu_3O_{7-\delta}$ [59] ($T_{C0}$ = 97 K [60]),[14] $YBa_2Cu_4O_8$ (optimized under hydrostatic pressure of 12 GPa) [61-63],[15] the mixed-valence Tl-based cuprate compounds [64-70],[16] $HgBa_2Ca_2Cu_3O_{8+\delta}$ ($\delta = 0.27 \pm 0.04$) at ambient pressure [71] and at 25 GPa [72][17] ($T_{C0}$ estimated at 145 K [73]). Under hydrostatic pressure, the 2D carrier density increases with decreasing unit cell volume. However, for materials which are non-optimal at ambient pressures, such as $YBa_2Cu_4O_8$, charge transfer can also occur so as to optimize the 2D hole and electron densities.

We know of at least two optimal compounds for which $\sigma > \sigma_0$, $HgBa_2CuO_{4.15}$ [74] and $HgBa_2CaCu_2O_{6.22}$ [75][18], $T_{C0}$ = 95 K and 127 K, respectively [71], where the two electrons per excess oxygen (0.15 and 0.22, respectively) are equally distributed among the $(3+\eta)$ layers (this is equivalent to dividing the carriers between the two reservoirs, and then by the average number of layers per reservoir) such that $\sigma = \sigma_0 + 2(0.15)/(3+1) = 0.3030$ and $\sigma = \sigma_0 + 2(0.22)/(3+2) = 0.3160$, respectively. We attribute this to the unique structure of the Hg compounds which provides vacancy locations for excess oxygen and the $\leq 2$ valence of Hg.

### *2.3. Calculating σ for other high-$T_C$ materials*

For the remaining high-$T_C$ compounds discussed herein, $\sigma$ is calculated by several methods. In cases where the charge content is determined by doping, $\sigma$ is obtained from the definition,

$$\sigma = \gamma\, [x - x_0]\,, \tag{2.4a}$$

where $[x - x_0]$ is the doping factor in which x denotes the charge content and $x_0$ is the minimum value of x required for superconductivity. The factor $\gamma$ modifies the effect of the dopant by considering a given compound's structure and/or charge state. For those compounds where $[x - x_0]$ cannot determined independently through doping, $\sigma$ can be calculated by scaling to $\sigma_0$ for $YBa_2Cu_3O_{6.92}$ according to,

$$\sigma = \gamma\, \sigma_0, \tag{2.4b}$$

where $\gamma$ is determined by analogous considerations as in equation (2.4a); for example, the calculation in section 2.2 of $\sigma$ for $YBa_2Cu_3O_{6.60}$ corresponds to $\gamma = 0.439$. For the unique case of the Hg-based compounds, $HgBa_2CuO_{4.15}$ and $HgBa_2CaCu_2O_{6.22}$, one can assume $\sigma = \sigma_0$ for $\delta = 0$ such that $\gamma = 1 + (2\delta/[3+\eta])/\sigma_0$, which evaluates to 1.3289 and 1.3860 for $\delta = 0.15$ and 0.22, respectively.

---

[13] Structure: table II of [55], x = 6.60; $\zeta$ defined as for [55]; $x_0$ same as [57].

[14] Structure: table VII of [59]; $\zeta$ measured between $Ba_z$ and average of $O(4)_z$ and $O(5)_z$.

[15] Lattice parameters at 12 GPa extrapolated from 0 GPa values ($a_0$ = 3.8413 Å, $b_0$ = 3.8708 Å, $c_0$ = 27.24 Å) assuming pressure derivatives in table 1 of [61]; $\zeta$ measured between $Ba_z$ and average of $O(2)_z$ and $O(3)_z$, calculated from table 1 derivatives [62]; $T_{C0}$ extracted using midpoint of 0-T resistance data (figure 2 of [63]).

[16] $T_{C0}$ for $Tl_2Ba_2Ca_2Cu_3O_{10}$ from [67]; $T_{C0}$ for $TlBa_2Ca_2Cu_3O_{9+\delta}$ from [70].

[17] For consistency assume 0 GPa lattice parameters, $a_0$ and $c_0$ from [71]. Lattice parameters at 25 GPa are extrapolated from figure 2 of [72]; $a/a_0 = 0.96$ and $c/c_0 = 0.91$); $\zeta$ measured between $Ba_z$ and $O(2)_z$ obtained by extrapolating theoretical curves in figure 7 of [72] to 25 GPa.

[18] Structure: tables 1 and 2 at 110 K in [75].

Consider, for example, a compound such as $La_{2-x}Sr_xCuO_4$ with a corresponding insulating undoped material. The absence of conductivity for $x = 0$ requires that the charge associated with Sr doping populate both the hole and electron reservoirs – equally. Also, since there are two SrO layers, the carriers would be distributed equally between them. Secondly, since the high-$T_C$ compounds are ionic in nature, one expects the available carrier density to be affected by the valences of the individual ions. Considering the relevant structural and electronic characteristics of the high-$T_C$ families and subfamilies, we have established that it is possible to calculate the 2D superconducting (and mediating) carrier density, either directly or relative to $YBa_2Cu_3O_{6.92}$, for the optimal compounds with $\gamma \neq 1$ by employing the following principles:

1. ***Charge Allocation Rules***
   a. Sharing between N (typically 2) ions or structural layers introduces a factor of 1/N in $\gamma$.
   b. The doping is shared equally between the hole and electron reservoirs, resulting in a factor of 1/2.
2. ***Valence Scaling Rules***
   a. Heterovalent substitution of a valence +3 ion mapped to a valence +2 ion corresponding to the $YBa_2Cu_3O_{7-\delta}$ structural type introduces a factor of 1/2 in $\gamma$.
   b. The factor $\gamma$ scales with the +2 (–2) cation (anion) structural stoichiometry associated with participating charge.
   c. The factor $\gamma$ scales with the net valence of the undoped mediating layer.

These rules are applied in absolute terms of cation (or anion) substitution or in relative terms by scaling with respect to $YBa_2Cu_3O_{6.92}$ (i.e. through application of rules 2.a, 2.b and 2.c), making it possible to calculate $\sigma$ for a variety of very diverse high-$T_C$ compounds. Herein, rule 2.a is applied to compounds where the +1 relative valence difference is associated with the type I central-layer cation(s), and rule 2.b generalizes to both cations and anions the application of relative scaling already used to calculate $\sigma$ for the "60 K" $YBa_2Cu_3O_{6.60}$ phase (where $\gamma = 0.439$). Bear in mind that $\sigma$ corresponds to the dominant superconducting carrier density: In the cuprates and organics, for example, $\sigma$ corresponds the superconducting hole condensate.

*2.3.1. Direct cation substitution of known doping (rules 1.a and 1.b)*

For $La_{1.837}Sr_{0.163}CuO_{4-\delta}$ [78][19] and $La_{1.8}Sr_{0.2}CaCu_2O_{6\pm\delta}$ [78][20] the charge in the two La/SrO layers and the $CuO_2$ layer(s) is determined by the $Sr^{+2}$ content ($x = 0.163$ and 0.2, respectively). Given that these two structures contain two La/SrO layers ($N = \nu = 2$) (rule 1.a), and recognizing that the added charge must also dope the $CuO_2$ layer(s) (rule 1.b), $\gamma = (1/2)(1/2) = 0.25$. From equation (2.4a) one obtains $\sigma$ in absolute terms as,

[19] Structure: table I of [76] for $x = 0.1625$ at 10 K; $T_{C0}$ estimated from [76] and [77]; optimal $x = 0.163$ from [77].

[20] Structure: table I of [78]; $T_{C0}$, x and $x_0$ from [79].

$$\sigma = \gamma\,[x - x_0] = 0.25\,[x - x_0]\,. \qquad (2.5a)$$

Thus, for $La_{1.837}Sr_{0.163}CuO_{4-\delta}$ and $La_{1.8}Sr_{0.2}CaCu_2O_{6\pm\delta}$, equation (2.5a) yields $\sigma = 0.0408$ and $\sigma = 0.05$, respectively (experimentally, $x_0 = 0$ [76,78]).

The n-type infinite layer compound, $(Sr_{0.9}La_{0.1})CuO_2$ [80],[21] is similarly straight forward. Since the added charge dopes both the hole and electron single layers ($\eta = \nu = 1$), $\gamma = 1/2$, and given that $x_0 = 0$ [81], we have,

$$\sigma = \gamma\,[x - x_0] = 0.5\,[0.1 - 0.0] = 0.05\,. \qquad (2.5b)$$

The ruthenate compounds $A_2YRu_{1-x}Cu_xO_6$ (with A = Sr or Ba; x = 0.05-0.15) are double-perovskites containing no cuprate planes [82]. For $Ba_2YRu_{0.9}Cu_{0.1}O_6$ ($T_{C0}$ ~ 30-40 K), $\eta = \nu = 1$, $\gamma = 1/2$ (rule 1.b) and $x_0 = 0$ such that,

$$\sigma = \gamma\,[x - x_0] = 0.5\,[0.1 - 0.0] = 0.05\,. \qquad (2.5c)$$

The ruthenate compounds $A_2YRu_{1-x}Cu_xO_6$ (with A = Ba or Sr; x = 0.05 – 0.15) are double-perovskites containing no cuprate planes and with $\eta = \nu = 1$ [81]. The determination of $\gamma$ follows from equation (2.5b), wherein rule 1b introduces the factor 1/2. In the lower limit, one expects a minimum of ~2 charges per Cu dopant, which are shared between two charge reservoirs of each layer type (AO, 1/2 $(YRu_{1-x}Cu_xO_4)$), producing a net factor of unity. Thus, for $Ba_2YRu_{0.9}Cu_{0.1}O_6$ (with $T_{C0}$ ~ 30–40 K), we estimate $\gamma = (1/2)(1) = 1/2$, yielding $\sigma = 0.05$ as stated by equation (2.5c).

While one may expect an average effective charge state for Ru near +5, and that of Cu to be between +2 and +3 (post anneal) [81a], the lower-limit estimation provided, which places the corresponding data point in figure 2 to the left of the line, appears sufficient to include the ruthenates with the other high-$T_C$ compounds found to follow equation (2.6) so far. Owing to the uncertainty in the experimental values for $T_{C0}$, as well as the Ru and Cu valance states, however, this compound was excluded in the data analyses presented below. Future research will attempt a more accurate determination of the charge per doped Cu, and thus $\sigma$.

*2.3.2. Cation substitution of doping layers (rules 1.a, 2.a and 2.b)*

For the compounds discussed in this section the absolute value of the participating charge associated with the dopant is not known, but by following the appropriate rules of comparative doping one can determine a value relative to $\sigma_0$. For the mixed-valence Bi/Pb–based compounds, $(Pb_{0.5}Cu_{0.5})Sr_2(Y_{0.6}Ca_{0.4})Cu_2O_{7-\delta}$ [83],[22] $Bi_2Sr_2CaCu_2O_{8+\delta}$ [85], $Bi_2Sr_2Ca_2Cu_3O_{10+\delta}$ [86],[23] $Pb_2Sr_2(Y_{0.5}Ca_{0.5})Cu_3O_8$ [88] ($T_{C0}$ = 75 K [89]) and $Bi_2(Sr_{1.6}La_{0.4})CuO_{6+\delta}$ [90],[24] the substitution of Bi or Pb for $Cu^{+2}$ in the central (or inner) type I reservoir layer(s) depletes the charge content.

In the case of $(Pb_{0.5}Cu_{0.5})Sr_2(Y_{0.6}Ca_{0.4})Cu_2O_{7-\delta}$ [83], the relative charge of the inner Pb/CuO layer is given by the fractions of $Cu^{+2}$ and $Pb^{+3}$ (formal valences +4 and +2; we assume an effective average valence of +3 for Pb based upon Pb-O bond-length analysis of the $[(Pb,Cu)O_x]$ layers [84]), weighted by their respective valences according to rule 2.a, yielding a relative scaling factor of $(0.5_{Pb}/2 + 0.5_{Cu})/1_{Cu}$, where the $1_{Cu}$ in the denominator corresponds to the full $Cu^{+2}$ in the chain layers of $YBa_2Cu_3O_{6.92}$ (rule 2.b) Given that the doping of the inner Pb/CuO layer is shared equally between the two outer SrO layers, rule 1a provides that $\gamma = 0.5\ (0.5/2 + 0.5) = 0.75$. From equation (2.4b) one obtains $\sigma$ as,

$$\sigma = \gamma\ \sigma_0 = 0.5\ (0.75)\ \sigma_0 = 0.0855\ . \qquad (2.5d)$$

---

[21] Structure: table 1 of [80]; x and $x_0$ from [81].

[22] *A*, *d*: table I of [83], sample 7; $\zeta$ from Table III of [84], site compositions 0.63 Pb; 0.37 Cu, and 0.85 Y; 0.15 Ca.

[23] Structure: tables 1 and 2 of [86]; $T_{C0}$ = 112 K [87].

[24] Structure: table 2 of [90]; $T_{C0}$ = 34 K [91].

The next three cases correspond to 100% substitution of $Bi^{+3}$ (or $Pb^{+3}$ in the third case; the +3 valence on the lead ion assumes $Pb^{+2}$ plus one half of central $Cu^{+2}$ ion charge) for $Cu^{+2}$ in the type I reservoir central layer, and with a doubling of these layers. Thus according to rule 2.a, one BiO (PbO) layer corresponds to a factor of $(1.0_{Bi/Pb}/2 + 0.0_{Cu})/1_{Cu} = 0.5$. As in equation (2.5*d*), sharing between the two SrO layers (rule 1.a) yields an additional factor of 1/2. Finally, given that there are two BiO (PbO) layers replacing one CuO layer (rule 2.b), $\gamma = (1.0_{Bi/Pb}/2 + 0.0_{Cu})(1/2)(2_{Bi/PbO}/1_{CuO}) = 0.5$ and,

$$\sigma = \gamma\ \sigma_0 = (0.5)(0.5)(2)\ \sigma_0 = 0.5\ \sigma_0 = 0.114\ . \qquad (2.5e)$$

For the single-layer material, $Bi_2(Sr_{1.6}La_{0.4})CuO_{6+\delta}$ [90], the 1/2 $\gamma$-factor (relative to $YBa_2Cu_3O_{6.92}$) in equation (2.5e) arising from the double BiO layer structure, would naturally apply. However, since $YBa_2Cu_3O_{7-\delta}$ has two corresponding $Ba^{+2}$ ions, the partial $Sr^{+2}$ doping (x = 1.6) of the outer layers introduces a relative doping factor of (1.6 – 1.16)/2 = 0.22 for the participating charge (rule 2.b), where $x_0 = 1.16$ [92], yielding

$$\sigma = \gamma\ \sigma_0 = 0.5\ (0.22)\ \sigma_0 = 0.0251\ . \qquad (2.5f)$$

In the case of the rutheno-cuprate compound, $RuSr_2GdCu_2O_8$ [93],[25] the structure contains a type I reservoir $SrO\text{-}RuO_2\text{-}SrO$, where the $Cu^{+2}O$ chain layer is replaced by a $Ru^{+5}O_2$ layer, and $Y^{+3}$ is replaced by $Gd^{+3}$. Taking the Ru charge state to be +5 in this material [94], and given the charge equivalence of $Bi^{+3}O^{-2}$ and $Ru^{+5}(O^{-2})_2$, one can draw an analogy with the Bi/Pb compounds of equation (2.5e), and approximate $\sigma$. In this case, however, there is only one layer that is charge equivalent to $Bi^{+3}O^{-2}$ yielding $\gamma = (1/2)(1/2) = 0.25$ and,

$$\sigma = \gamma\ \sigma_0 = 0.25\ \sigma_0 = 0.0570\ . \qquad (2.5g)$$

### *2.3.3. Iron pnictides: anion/cation substitution of known doping (rules 1.a and 1.b)*

In the (n-type) Ln-O/F-Fe-As ("1111") iron pnictides, e.g. $La(O_{0.92-y}F_{0.08})FeAs$ [95], $Ce(O_{0.84-y}F_{0.16})FeAs$ [96], $Tb(O_{0.80-y}F_{0.20})FeAs$ [97][26] and $Sm(O_{0.65-y}F_{0.35})FeAs$ [98][27] (where y accounts for the actual O-site occupancy; we assume $[x - x_0]$ is given by F stoichiometry), the Coulombic interaction is assumed to occur between the Ln(O/F) and AsFe layers, which defines $\zeta$ (Ln-As distance along the *c*-axis), and sets $\eta = \nu = 1$ (i.e. within the periodicity *d*). Applying rules 1.b (equally shared doping between the hole and electron reservoirs) and 1.a (where the doping is further divided between component layers of the two reservoirs, {O/F, Ln} and {As, Fe}), one obtains $\gamma = (1/4)$ and,

$$\sigma = \gamma\ [x - x_0] = 0.25\ [x - x_0]\ , \qquad (2.5h)$$

where x is the fluorine content and $x_0 = 0$. For the $Th^{+4}$ doped n-type 1111 iron-pnictide compounds, such as $(Sm_{0.7}Th_{0.3})OFeAs$ [100], the value of $\sigma$ is also given by equation (2.5h), yielding $\sigma = 0.075$. The symmetry of the 1111 compound structure makes it is unnecessary to assign reservoir types; tentatively we set the FeAs structures as type I and the Ln(O/F) as type II.

---

[25] Structure: table 1 of [93]; Assume $T_{C0} = 50$ K because transition is broadened due to Ru and Gd moments.

[26] Structure: table 1 of [97]; *A* and *d* for x = 0.2 (sharpest resistive transition); $\zeta$ from figure 1 caption of [97] (x = 0.1).

[27] Structure: table 1 of [98]; $T_{C0} = 55$ K [99].

For the Ba-$Fe_2$-$As_2$ ("122") p-type compound, $(Ba_{0.6}K_{0.4})Fe_2As_2$ ($T_{C0}$ = 37 K, x = 0.4, $x_0$ = 0) [101],[28] the Coulombic interaction occurs between the Ba/K layer and the two adjacent $Fe^{+2}As^{-3}$ structures. As above, rules 1.a and 1.b apply, but in this case the doping is also shared between the two FeAs layer structures (rule 1.a) yielding an additional factor of 1/2 in γ such that,

$$\sigma = \gamma\,[x - x_0] = 0.125\,[0.4] = 0.05\,, \tag{2.5i}$$

where γ = (1/8) and $x_0$ = 0. The σ parameter for the n-type analogue to the above compound, $Ba(Fe_{1.84}Co_{0.16})As_2$ [102], is similarly calculated as in equation (2.5i), yielding σ = 0.02. In calculating equation (2.5i) we designate the Ba(K) layers as type II (η = 1), which interact with the As in the FeAs (type I, ν = 2) structures.

*2.3.4. Charge-transfer salts (rules 1.a and 2.c)*

For κ–$[BEDT–TTF]_2Cu[N(CN)_2]Br$ [103], the hole conduction is in the ac-plane along the sulfur chains of the two BEDT-TTF molecules (each of which is bisected by a centrally located C–C bond). Equating the $Cu^{+1}[N(CN)_2]^{-1}Br^{-1}$ anion layer with $Cu^{+2}[O^{-2}][O^{-2}]$ (type II layer) shows a factor of 1/2 between the valences of the cuprate plane ions compared to those comprising the $Cu[N(CN)_2]Br$ anion molecule, leading to a base anion layer charge of $\sigma_0/2$ (rule 2.c), which must equal the positive charge available to the two BEDT-TTF (type I layer) molecules. Dividing this charge between the two BEDT-TTF molecules comprising the dimer (rule 1.a), and further distributing the charge among the two halves of the BEDT-TTF molecule (rule 1.a) yields an additional factor of (1/2)(1/2) such that γ = (1/2)(1/2)(1/2) = 0.125. From equation (2.4b) one obtains,

$$\sigma = \gamma\,\sigma_0 = 0.125\,\sigma_0 = 0.0285. \tag{2.5j}$$

A reasonable value for ζ is determined from the nearest-neighbor S–$Cu[N(CN)_2]Br$ distance.[29]

### *2.4. Comparison with experiment*

We have established that the high-$T_C$ pairing mechanism is Coulombic in nature, formulated a set of parameters and equations defining our model, and developed a method of determining the 2D carrier density in various high-$T_C$ families. Table 1 provides compiled data for $T_{C0}$, σ, ζ, *A*, *d*, η, ν, and γ, along with the structures comprising the type I and type II charge reservoirs, corresponding to the 31 optimum (or near optimum) superconducting compounds above. The components of the γ factor for compounds with γ ≠ 1 are given in table 2. Owing to doping-induced inhomogeneity, the measured properties of non-optimal materials are unrepresentative of the intrinsic superconducting state and thus not included. Validation of our $n_{2D}$ values can be found by considering the reference compound $YBa_2Cu_3O_{6.92}$, where the parameters in table 1 yield a volume hole carrier density of $n_{3D} = \nu\sigma/dA = 2.63 \times 10^{21}$ cm$^{-3}$, which is in good agreement with the experimental value of $n_{3D} = 2.40 \pm 0.05 \times 10^{21}$ cm$^{-3}$ obtained from Hall coefficient and resistivity in the normal state of bulk-crystals of $YBa_2Cu_3O_{7-\delta}$, [19]; it is also consistent with electrostatic charging experiments on thin films [42].

---

[28] Structure: table I of [101] at 20 K.

[29] Structure: table I of [103] (*A*, *d* obtained at 9 K); ζ estimated assuming the interacting holes are primarily located in the sulfurs nearest to the anion layers; ζ ≈ (1/6)[*b*/2] Å = 2.4579 Å. Also, *d* = *b*/2 = 14.7475 Å and *A* = *ac*/2 = 54.4745 Å$^2$ (Z = 4); $T_{C0}$ from [104] (inductive onset at 11.6 K and zero resistance at 10.5 K).

**Table 1.** Transition temperature $T_{C0}$, electronic ($\sigma$, $\gamma$, P/N) and structural ($\zeta$, $A$, $d$, $\eta$, $\nu$) parameters for 31 optimal high-$T_C$ superconductors discussed in section 2. The first two columns list the reference number(s) and compound. The last two columns identify the dominant superconducting carrier type and type I / type II reservoirs per formula unit ($O_x$ denotes partial filling). Related compound families are grouped accordingly.

| Refs. | Superconducting Compound | $T_{C0}$ (K) | $\sigma$ (Type I) | $\zeta$ (Å) | $A$ (Å$^2$) | $d$ (Å) | $\eta$ | $\nu$ | $\gamma$ | P/N | Type I Reservoirs<br>Type II Reservoirs |
|---|---|---|---|---|---|---|---|---|---|---|---|
| 55-57 | $YBa_2Cu_3O_{6.92}$ | 93.7 | $\sigma_0$ | 2.2677 | 14.8596 | 11.6802 | 2 | 2 | 1 | P | BaO-CuO-BaO<br>$CuO_2$-Y-$CuO_2$ |
| 55,58 | $YBa_2Cu_3O_{6.60}$ | 63 | 0.439 $\sigma_0$ | 2.2324 | 14.8990 | 11.7279 | 2 | 2 | 0.439 | P | BaO-CuO-BaO<br>$CuO_2$-Y-$CuO_2$ |
| 59,60 | $LaBa_2Cu_3O_{7-\delta}$ | 97 | $\sigma_0$ | 2.1952 | 15.3306 | 11.8180 | 2 | 2 | 1 | P | BaO-CuO-BaO<br>$CuO_2$-La-$CuO_2$ |
| 61-63 | $YBa_2Cu_4O_8$ (12 GPa) | 104 | $\sigma_0$ | 2.1658 | 14.2060 | 12.9042 | 2 | 2 | 1 | P | BaO-CuO-CuO-BaO<br>$CuO_2$-Y-$CuO_2$ |
| 64 | $Tl_2Ba_2CuO_6$ | 80 | $\sigma_0$ | 1.9291 | 14.9460 | 11.6195 | 1 | 2 | 1 | P | BaO-TlO-TlO-BaO<br>$CuO_2$ |
| 65 | $Tl_2Ba_2CaCu_2O_8$ | 110 | $\sigma_0$ | 2.0139 | 14.8610 | 14.6590 | 2 | 2 | 1 | P | BaO-TlO-TlO-BaO<br>$CuO_2$-Ca-$CuO_2$ |
| 66,67 | $Tl_2Ba_2Ca_2Cu_3O_{10}$ | 130 | $\sigma_0$ | 2.0559 | 14.8248 | 17.9400 | 3 | 2 | 1 | P | BaO-TlO-TlO-BaO<br>$CuO_2$-Ca-$CuO_2$-Ca-$CuO_2$ |
| 68 | $TlBa_2CaCu_2O_{7-\delta}$ | 103 | $\sigma_0$ | 2.0815 | 14.8734 | 12.7540 | 2 | 2 | 1 | P | BaO-TlO-BaO<br>$CuO_2$-Ca-$CuO_2$ |
| 69,70 | $TlBa_2Ca_2Cu_3O_{9+\delta}$ | 133.5 | $\sigma_0$ | 2.0315 | 14.7686 | 15.8710 | 3 | 2 | 1 | P | BaO-TlO-BaO<br>$CuO_2$-Ca-$CuO_2$-Ca-$CuO_2$ |
| 71 | $HgBa_2Ca_2Cu_3O_{8+\delta}$ | 135 | $\sigma_0$ | 1.9959 | 14.8060 | 15.7782 | 3 | 2 | 1 | P | BaO-$HgO_x$-BaO<br>$CuO_2$-Ca-$CuO_2$-Ca-$CuO_2$ |
| 72,73 | $HgBa_2Ca_2Cu_3O_{8+\delta}$ (25 GPa) | 145 | $\sigma_0$ | 1.9326 | 13.6449 | 14.3582 | 3 | 2 | 1 | P | BaO-$HgO_x$-BaO<br>$CuO_2$-Ca-$CuO_2$-Ca-$CuO_2$ |
| 74,71 | $HgBa_2CuO_{4.15}$ | 95 | $\sigma_0$+0.075 | 1.9214 | 15.0362 | 9.5073 | 1 | 2 | 1.3289 | P | BaO-$HgO_x$-BaO<br>$CuO_2$ |
| 75,71 | $HgBa_2CaCu_2O_{6.22}$ | 127 | $\sigma_0$+0.088 | 2.039 | 14.9375 | 12.230 | 2 | 2 | 1.3860 | P | BaO-$HgO_x$-BaO<br>$CuO_2$-Ca-$CuO_2$ |
| 76,77 | $La_{1.837}Sr_{0.163}CuO_{4-\delta}$ | 38 | 0.0408 | 1.7828 | 14.2268 | 6.6029 | 1 | 2 | 0.25 | P | La/SrO-La/SrO<br>$CuO_2$ |
| 78,79 | $La_{1.8}Sr_{0.2}CaCu_2O_{6\pm\delta}$ | 58 | 0.05 | 1.7829 | 14.3761 | 9.6218 | 2 | 2 | 0.25 | P | La/SrO-La/SrO<br>$CuO_2$-Ca-$CuO_2$ |
| 80,81 | $(Sr_{0.9}La_{0.1})CuO_2$ | 43 | 0.05 | 1.7051 | 15.6058 | 3.4102 | 1 | 1 | 0.50 | N | Sr/La<br>$CuO_2$ |
| 82 | $Ba_2YRu_{0.9}Cu_{0.1}O_6$ | 35 | 0.05 | 2.0809 | 17.3208 | 4.1618 | 1 | 1 | 0.50 | P | BaO<br>½($YRu_{0.9}Cu_{0.1}O_4$) |
| 83,84 | $(Pb_{0.5}Cu_{0.5.})Sr_2(Y,Ca)Cu_2O_{7-\delta}$ | 67 | 0.375 $\sigma_0$ | 1.9967 | 14.5771 | 11.8290 | 2 | 2 | 0.375 | P | SrO-Pb/CuO-SrO<br>$CuO_2$-Y/Ca-$CuO_2$ |
| 85 | $Bi_2Sr_2CaCu_2O_{8+\delta}$ (unannealed) | 89 | 0.5 $\sigma_0$ | 1.795 | 14.6665 | 15.4450 | 2 | 2 | 0.50 | P | SrO-BiO-BiO-SrO<br>$CuO_2$-Ca-$CuO_2$ |
| 86,87 | $(Bi,Pb)_2Sr_2Ca_2Cu_3O_{10+\delta}$ | 112 | 0.5 $\sigma_0$ | 1.6872 | 14.6340 | 18.5410 | 3 | 2 | 0.50 | P | SrO-BiO-BiO-SrO<br>$CuO_2$-Ca-$CuO_2$-Ca-$CuO_2$ |
| 88,89 | $Pb_2Sr_2(Y,Ca)Cu_3O_8$ | 75 | 0.5 $\sigma_0$ | 2.028 | 14.6458 | 15.7334 | 2 | 2 | 0.50 | P | SrO-PbO-Cu-PbO-SrO<br>$CuO_2$-Y/Ca-$CuO_2$ |
| 90-92 | $Bi_2(Sr_{1.6}La_{0.4})CuO_{6+\delta}$ | 34 | 0 .11 $\sigma_0$ | 1.488 | 14.5422 | 12.1995 | 1 | 2 | 0.11 | P | SrO-BiO-BiO-SrO<br>$CuO_2$ |
| 93,94 | $RuSr_2GdCu_2O_8$ | 50 | 0.25 $\sigma_0$ | 2.182 | 14.7372 | 11.5652 | 2 | 2 | 0.25 | P | SrO-$RuO_2$-SrO<br>$CuO_2$-Gd-$CuO_2$ |
| 95 | $La(O_{0.92-y}F_{0.08})FeAs$ | 26 | 0.02 | 1.7677 | 16.1620 | 4.3517 | 1 | 1 | 0.25 | N | ½(As-2Fe-As)<br>½(La-2O/F-La) |
| 96 | $Ce(O_{0.84-y}F_{0.16})FeAs$ | 35 | 0.04 | 1.6819 | 15.8778 | 4.3016 | 1 | 1 | 0.25 | N | ½(As-2Fe-As)<br>½(Ce-2O/F-Ce) |
| 97 | $Tb(O_{0.80-y}F_{0.20})FeAs$ | 45 | 0.05 | 1.5822 | 14.8996 | 4.1660 | 1 | 1 | 0.25 | N | ½(As-2Fe-As)<br>½(Tb-2O/F-Tb) |
| 98,99 | $Sm(O_{0.65-y}F_{0.35})FeAs$ | 55 | 0.0875 | 1.667 | 15.4535 | 4.2328 | 1 | 1 | 0.25 | N | ½(As-2Fe-As)<br>½(Sm-2O/F-Sm) |
| 100 | $(Sm_{0.7}Th_{0.3})OFeAs$ | 51.5 | 0.075 | 1.671 | 15.4897 | 4.2164 | 1 | 1 | 0.25 | N | ½(As-2Fe-As)<br>½(Sm/Th-2O/F-Sm/Th) |
| 101 | $(Ba_{0.6}K_{0.4})Fe_2As_2$ | 37 | 0.05 | 1.932 | 15.2803 | 6.6061 | 1 | 2 | 0.125 | P | As-2Fe-As<br>Ba/K |
| 102 | $Ba(Fe_{1.84}Co_{0.16})As_2$ | 22 | 0.02 | 1.892 | 15.6848 | 6.4897 | 1 | 2 | 0.125 | N | As-2Fe/Co-As<br>Ba |
| 103,104 | $\kappa$–[BEDT-TTF]$_2$Cu[N(CN)$_2$]Br | 10.5 | 0.125 $\sigma_0$ | 2.4579 | 54.4745 | 14.7475 | 1 | 2 | 0.125 | P | S-chains [BEDT-TTF]$_2$<br>Cu[N(CN)$_2$]Br |

**Table 2.** Calculating the factor $\gamma$ of equations (2.4a) and (2.4b) using the rules defined in section 2.3. The first two columns provide the reference number(s) and name of the compound. The following columns give shared doping factors (rules 1.a and 1.b); type I reservoir relative number of outer and inner layers, and comparative valence differences (with respect to $YBa_2Cu_3O_{7-\delta}$, rule 2.a); relative doping factors (rule 2.b); type II reservoir comparative valence differences (rule 2.c); and $\gamma$. Related compound families are grouped accordingly. Entries are shown where relevant.

| Refs. | Superconducting Compound | Shared Doping | | Type I | | | Relative Doping (rule 2.b) | Type II | $\gamma$ |
|---|---|---|---|---|---|---|---|---|---|
| | | intra-type (rule 1.a) | inter-type (rule 1.b) | rel. no. outer | rel. no. inner | Δ valence (rule 2.a) | | Δ valence (rule 2.c) | |
| 55,56 | $YBa_2Cu_3O_{6.60}$ | - | - | 1 | 1 | - | 0.439 | - | 0.439 |
| 74,71 | $HgBa_2CuO_{4.15}$ | - | - | 1 | 1 | - | 1.3289 [a] | - | 1.3289 |
| 75,71 | $HgBa_2CaCu_2O_{6.22}$ | - | - | 1 | 1 | - | 1.3860 [a] | - | 1.3860 |
| 76,77 | $La_{1.837}Sr_{0.163}CuO_{4-\delta}$ | 1/2 | 1/2 | - | - | - | - | - | 1/4 |
| 78,79 | $La_{1.8}Sr_{0.2}CaCu_2O_{6\pm\delta}$ | 1/2 | 1/2 | - | - | - | - | - | 1/4 |
| 80,81 | $(Sr_{0.9}La_{0.1})CuO_2$ | - | 1/2 | - | - | - | - | - | 1/2 |
| 82 | $Ba_2YRu_{0.9}Cu_{0.1}O_6$ | - | 1/2 | - | - | - | - | - | 1/2 |
| 83,84 | $(Pb_{0.5}Cu_{0.5.})Sr_2(Y,Ca)Cu_2O_{7-\delta}$ | 1/2 | - | 1 | 1 | (0.25+0.5) [b] | - | - | 3/8 |
| 85 | $Bi_2Sr_2CaCu_2O_{8+\delta}$ (unannealed) | 1/2 | - | 1 | 2 | 1/2 | - | - | 1/2 |
| 86,87 | $(Bi,Pb)_2Sr_2Ca_2Cu_3O_{10+\delta}$ | 1/2 | - | 1 | 2 | 1/2 | - | - | 1/2 |
| 88,89 | $Pb_2Sr_2(Y,Ca)Cu_3O_8$ | 1/2 | - | 1 | 2 | 1/2 | - | - | 1/2 |
| 90-92 | $Bi_2(Sr_{1.6}La_{0.4})CuO_{6+\delta}$ | 1/2 | - | 1 | 2 | 1/2 | 0.22 | - | 0.11 |
| 93,94 | $RuSr_2GdCu_2O_8$ | 1/2 | - | 1 | 1 | 1/2 | - | - | 1/4 |
| 95 | $La(O_{0.92-y}F_{0.08})FeAs$ | 1/2 | 1/2 | - | - | - | - | - | 1/4 |
| 96 | $Ce(O_{0.84-y}F_{0.16})FeAs$ | 1/2 | 1/2 | - | - | - | - | - | 1/4 |
| 97 | $Tb(O_{0.80-y}F_{0.2})FeAs$ | 1/2 | 1/2 | - | - | - | - | - | 1/4 |
| 98,99 | $Sm(O_{0.65-y}F_{0.35})FeAs$ | 1/2 | 1/2 | - | - | - | - | - | 1/4 |
| 100 | $(Sm_{0.7}Th_{0.3})OFeAs$ | 1/2 | 1/2 | - | - | - | - | - | 1/4 |
| 101 | $(Ba_{0.6}K_{0.4})Fe_2As_2$ | 1/2 • 1/2 | 1/2 | - | - | - | - | - | 1/8 |
| 102 | $Ba(Fe_{1.84}Co_{0.16})As_2$ | 1/2 • 1/2 | 1/2 | - | - | - | - | - | 1/8 |
| 103,104 | κ–[BEDT-TTF]$_2$Cu[N(CN)$_2$]Br | 1/2 • 1/2 | - | - | - | - | - | 1/2 | 1/8 |

[a] Calculations in section 2.3 assuming equation (2.4b).
[b] Corresponding to a 50% partial $Pb^{+3}$ substitution.

### *2.4.1. Calculating the optimal transition temperature $T_{C0}$*

Accepting that the pairing potential is Coulombic, and depends on the two lengths $\zeta$ and $\ell$ (calculated according to equation (2.3) from the parameters in table 1), one would expect the transition temperature to vary according to equation (2.2). A non-linear regression fit to the data of table 1 yields p = 1.02 ± 0.01 and q = 1.06 ± 0.04; thus for simplicity we take p = q = 1 such that we can write,

$$k_B T_{C0} = \beta(\sigma\, \eta/A)^{1/2} \,/\, \zeta = \beta\,/\ell\zeta\,. \qquad (2.6)$$

The validity of equation (2.6) is presented in figure 2 where we plot $T_{C0}$ versus $(\sigma\,\eta/A)^{1/2} / \zeta$ (experimental uncertainties in $T_{C0}$ are in the range ±(1–2) K, except for $Ba_2YRu_{0.9}Cu_{0.1}O_6$: ±5 K). As can be seen, the transition temperatures of the optimal compounds closely follow a straight line (linear correlation coefficient r = 0.9987) with slope 1247.4 ± 3.7 K Å$^2$ and zero intercept (to 0.6 K accuracy), as predicted by equation (2.6), yielding the universal constant, $\beta$ = 0.1075 ± 0.0003 eV Å$^2$, with a standard deviation in calculated $T_{C0}$ of 1.4 K. Our results are also consistent with equation (2.1), expressed as $\zeta/\ell < 1$; the plot of $\zeta/\ell$ against $T_{C0}$ provided in figure 3 shows (i) that for these materials, $\zeta$ is significantly less than $\ell$, as anticipated from equation (2.1), and (ii) that $T_{C0}$ tends to increase with increasing $\zeta/\ell$. Additionally, figure 4 presents graphs of the interlayer spacing $\zeta$ and $\ell$, each versus $T_{C0}$ (symbols in figures 2 – 4 correspond to the compounds given in the legend in figure 2). The dashed line in figure 4(a) is the average $\langle\zeta\rangle$ = 1.94 Å (some trends noted: $\zeta > \langle\zeta\rangle$ for Y-Ba-Cu-O compounds, $\zeta < \langle\zeta\rangle$ for "1111" Fe-As compounds); figure 4(b) shows that $T_{C0}$ is not a smooth function of $\ell$ (dotted line is trend $\ell \propto T_{C0}^{-1}$). Also, it is apparent that $\ell$ varies more strongly as compared to $\zeta$ among the high-$T_C$ compounds.

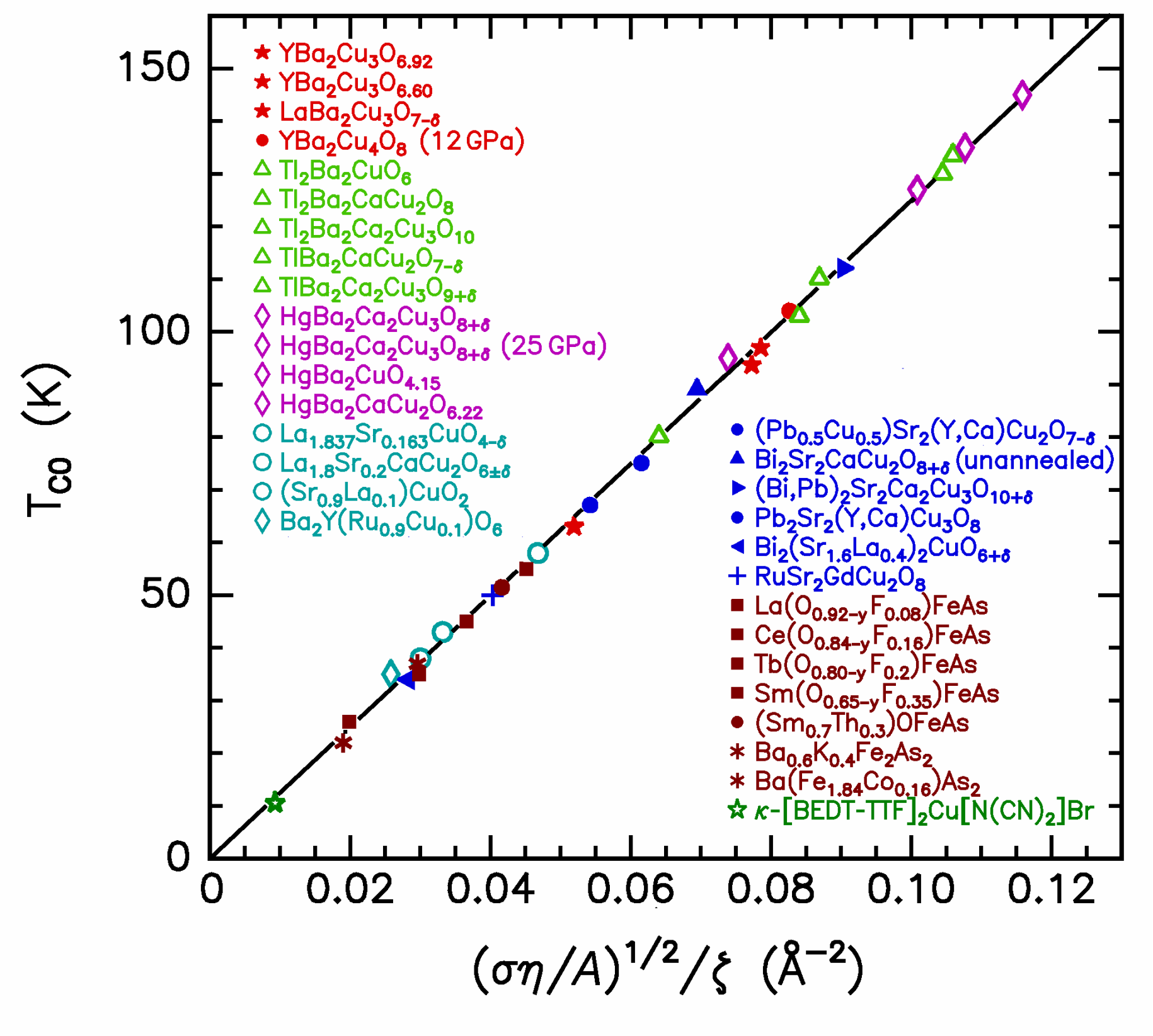

**Figure 2.** $T_{C0}$ versus $(\sigma\eta/A)^{1/2}/\zeta$ for the optimal compounds given in table 1. The solid line corresponds to the function given in equation (2.6). The compounds are grouped by families (colour) and distinguished by structures (symbols); legend list is in order of appearance in table 1.

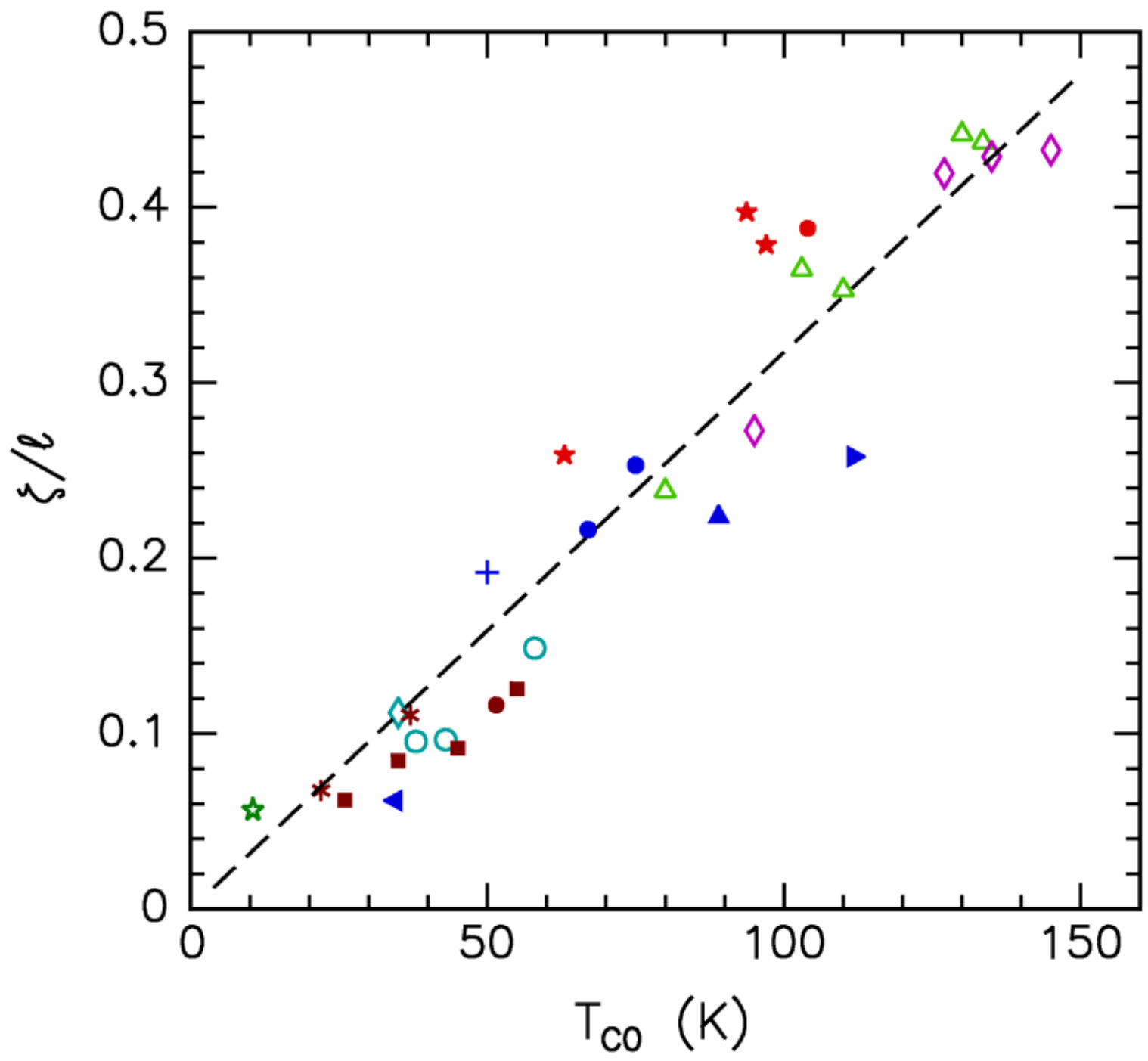

**Figure 3.** Graph of ζ/ℓ versus $T_{C0}$ (data from table 1); ζ is the interlayer spacing and ℓ is the mean intralayer distance between superconducting interaction charges. Dashed line denotes trend. Symbols correspond to legend in figure 2.

The verification of equation (2.6) provided in figure 2, along with previous research [5,6,19], proves that the high-$T_C$ pairing mechanism is Coulombic in nature, governed by the electrodynamic interactions between positive and negative charges separated by a physical distance ζ. Specifically, we identify the interaction to be between the carriers associated with the nearest-neighbor ions of the type I cation (e.g. BaO) and type II anion (e.g. $CuO_2$) layers. Note that for realistic $\zeta = \langle\zeta\rangle = 1.94$ Å, and taking the upper limit of equation (2.1) with $\ell = \zeta$, equation (2.6) predicts the existence of a "room temperature" superconductor with $T_{C0} = 331\pm1$ K.

*2.4.2 Interpreting the constant β*

The constant β is assumed to be the product of $e^2$ (e is the electron charge) corresponding to the Coulomb potential across ζ, and a fundamental constant Λ, which has units of length and is determined (to ± 0.3%) as,

$$\Lambda = \beta / e^2 = (0.1075 \text{ eV-Å}^2) / (14.4 \text{ eV-Å}) = 0.00747 \text{ Å} . \quad (2.7)$$

The possible presence of another numerical factor is subsumed into this definition of Λ (for this simple calculation we do not correct for the materials dielectric constant). Thus, equation (2.6) can be rewritten as,

$$k_B T_{C0} = e^2 \Lambda / \ell\zeta , \quad (2.8)$$

where the energy scale is set by the Coulomb energy, modified by the factor Λ/ℓ.

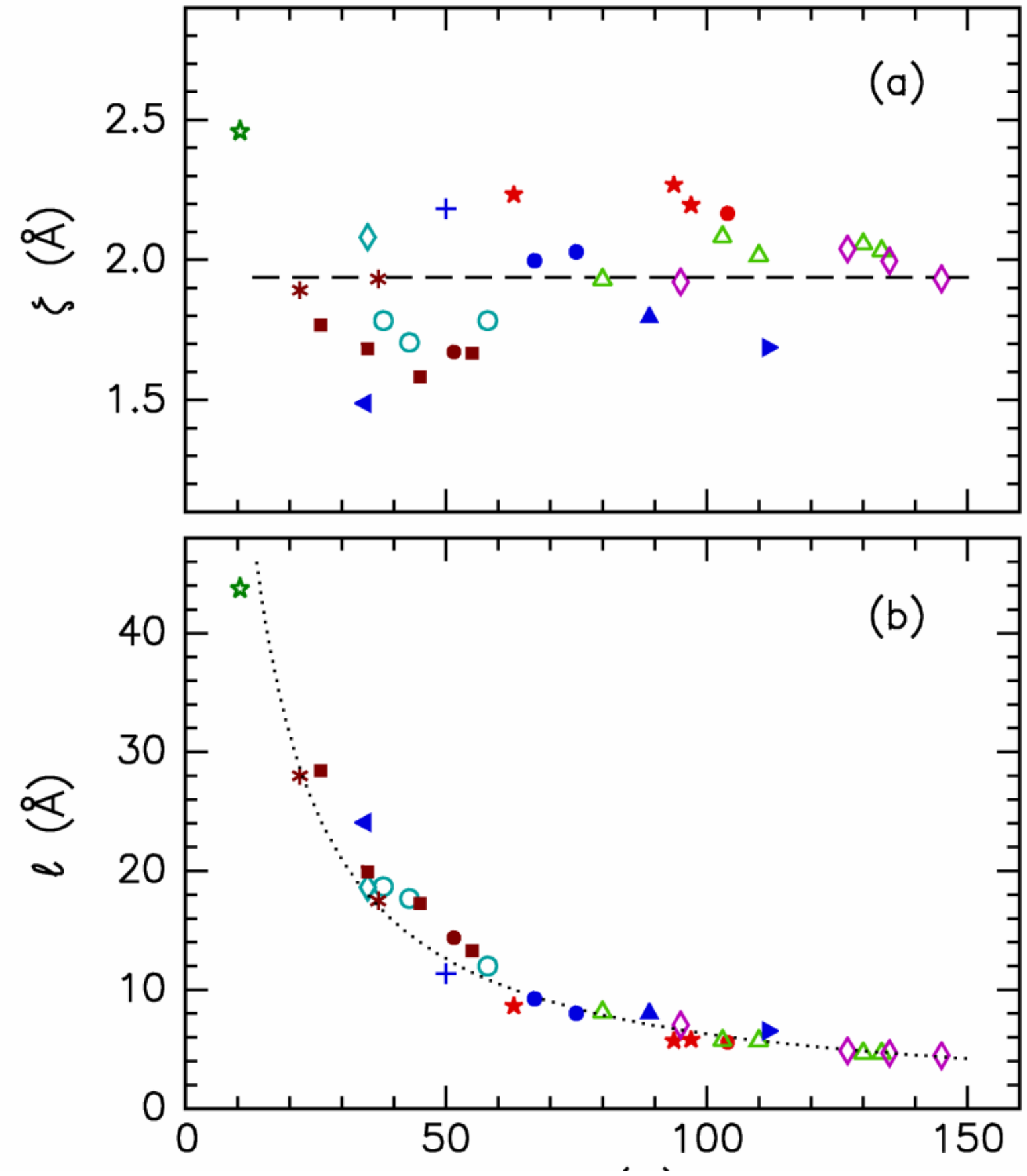

**Figure 4.** Graphs of the interlayer spacing ζ [frame (a)] and the mean intralayer spacing between the superconducting interaction charges ℓ [frame (b)], versus $T_{C0}$ as determined from data in table 1. Dashed line: average ζ; dotted curve: $\ell \propto T_{C0}^{-1}$. Symbols correspond to legend in figure 2.

## 3. Interlayer Coulomb interactions

The successful test of the Coulombic model presented in section 2 confirms that pairing involves two electronic bands that are spatially confined and physically separated from one another. The mechanism of superconductive coupling being applied in this work follows from the prescriptions for excitonic interactions presented by Little [44] and bears some similarity to the excitonic superconductivity discussed by Bardeen et al. [105]. In the general case, charges in only one of the layers need be treated as itinerant carriers (e.g. as in [44] or [106,107]). The form of equation (2.6) suggests pairing that is localized in real space, which invokes the premise of strong coupling theory [108], as distinguished from non-local pairing in BCS theory [3]. Justification for considering strong coupling for high-$T_C$ superconductors includes the large ratio of the superconducting energy gap to $k_BT_{C0}$ (on the order of 4 and about twice that of weak to intermediate coupling BCS superconductors). Another is non-BCS or two-fluid behaviour, where the gap persists to higher temperatures than for conventional metals [109].

Strong-coupling theory founded on the electron-phonon interaction provides guidance in understanding the origin of the algebraic (non-exponential) form for $T_{C0}$ in equation (2.6). Superconducting transition temperatures of strong phonon-coupled superconductors were calculated by McMillan [110]. The reanalysis by Allen and Dynes [111] showed that in the limit of large λ the transition temperature reduces to an algebraic form (see also [112] and [7]),

$$k_B T_{C0} = 0.182\,\hbar(\langle\omega^2\rangle\tilde{\lambda})^{1/2}, \qquad (3.1)$$

where $\hbar\langle\omega^2\rangle^{1/2}$ is an energy determined from the phonon spectrum and $\tilde{\lambda} = \lambda f(\mu^*)$ ($\lambda$ is the electron-phonon coupling parameter, $f(\mu^*) = (1+2.6\mu^*)^{-1}$, and $\mu^*$ is the Coulomb repulsion parameter). The product $\langle\omega^2\rangle\lambda$ involved in equation (3.1) is, apart from the factor $f(\mu^*)$, determined by the first moment of the electron-phonon spectral function $\alpha^2(\omega)\,F(\omega)$, where $\alpha^2(\omega)$ is the electron-phonon coupling and $F(\omega)$ is the density of phonon states:

$$\langle\omega^2\rangle\lambda = 2\int_0^\infty d\omega\,\omega\,\alpha^2(\omega)\,F(\omega) \quad . \qquad (3.2)$$

Determining $T_{C0}$ from $\langle\omega^2\rangle\lambda$ according to equation (3.1) turns out to be a good approximation when $\lambda$ is greater than about 2 [110] or 4 [111] because $T_C/(\langle\omega^2\rangle\lambda)^{½}$ approaches a constant at large $\lambda$. McMillan showed that the first moment integral in equation (3.2) is independent of phonon frequencies [110], because the electron-phonon matrix element is inversely proportional to the phonon frequency. This integral is obtained from the Coulomb pseudo-potential of the electron-ion interaction, v(q), where $q = k-k'$ connects electron states k and k′ on the Fermi surface, and it is proportional to the Fermi-surface average $\langle q^2v^2(q)\rangle_{FS}$ [110,113]. If the mediating boson is now taken to be an electronic excitation (of dispersion that need not be specified) the result, $\langle\omega^2\rangle\lambda \propto \langle q^2v^2(q)\rangle_{FS}$, can be utilized by taking the interlayer Coulomb potential for v(q).

Significantly, the two-dimensional Fermi contours for the optimally doped cuprate compounds are typically observed to be patches and arcs by probes such as angle resolved photoemission spectroscopy (ARPES) [40]. In the absence of sharp and closed Fermi contours the average of $q^2v^2(q)$ would need to be taken over a more extended region in q space. We proceed by making the intuitive ansatz of extending $\langle q^2v^2(q)\rangle_{FS}$ to an unrestricted average in q space. Consequently, the pairing interaction is not only local in real space (as in legacy strong-coupling theory) but it is also assumed to be nearly instantaneous and not time-retarded (as in the original BCS theory). By Fourier transformation, the average of $q^2v^2(q)$ in q-space is evaluated as the average of the interaction force in real space, i.e. we write

$$\left\langle q^2v^2(q)\right\rangle \propto \left\langle F^2(r)\right\rangle , \qquad (3.5)$$

where $F(r) = \pm\, e^2 r(r^2 + \zeta^2)^{-3/2}$ is the in-plane component of the Coulomb force of interaction between charges in planes separated by the distance $\zeta$ transverse to the planes and projected distance r parallel to the planes. As the centres of no ionic shells lie between the layers, we take unity for the (high-frequency) medium dielectric constant. Since the 2D density of interacting charges is $\ell^{-2}$, the average of $F^2(r)$ is thus obtained as

$$\left\langle F^2(r)\right\rangle = \ell^{-2}\int_0^\infty 2\pi r\, F^2(r)\, dr = \frac{\pi e^4}{2\ell^2\zeta^2} \quad . \qquad (3.6)$$

From equations (3.1) and (3.2) and this approach to finding $\langle\omega^2\rangle\lambda$, one directly obtains the result, $T_{C0} \propto e^2/\ell\zeta$. This confirms the expression for $T_{C0}$ given in equation (2.8) and in particular validates the presence of the factor $e^2$ as required for a Coulombic energy scale. Note that alternative theories scaling $T_C$ with plasmon energy yield a linear, rather than quadratic, dependence on e.

## 4. Discussion

Since their discovery, the list of high-$T_c$ superconductor families has grown to include (i) the (well-known) layered cuprate perovskites (e.g. $YBa_2Cu_3O_{7-\delta}$), (ii) rutheno-cuprates such as $GdSr_2Cu_2RuO_8$ [114], and certain Cu-doped layered ruthenates [83,115] such as $Ba_2YRu_{1-x}Cu_xO_6$ ($T_{C0}$ = 32.2 K, from equation (2.6), is well within the 30-40 K range given in [82]), (iii) selected 2D organic superconductors such as κ–$[BEDT\text{-}TTF]_2Cu[NCS]_2$ and κ–$[BEDT\text{-}TTF]_2Cu[N(CN)_2]Br$ [103], where the superconducting hole condensate resides in the sulfur chains of the BEDT-TTF molecules, (iv) the n-type 1111 iron pnictides, e.g. $Ln(O_{1-x}F_x)FeAs$ or $(Ln_{1-x}Th_x)FeAsO$, (v) the p-type 1111 iron-pnictides, $(Ln_{1-x}Ae_x)FeAsO$, where "Ae" denotes alkaline earth, (vi) the p-type 122 iron pnictides, e.g. $Ba_{0.6}K_{0.4}Fe_2As_2$, and possibly, (vii) the transuranium superconductors (e.g. $PuCoGa_5$). Since the latter five families (iii-vii) of compounds in addition to the ruthenates do not contain cuprate planes, it is manifestly clear that $CuO_2$ planes are not specifically required for high-$T_C$ superconductivity.

Several families of likely high-$T_C$ superconductors could not be included in the analysis presented herein for various reasons. Superconductors in the n-type T′-structure family $Ln_{2-x}Ce_xCuO_{4+\delta}$ (Ln = La, Pr, Nd, Sm, Eu), in which Hall effect and ARPES indicate separate electron and hole bands, were not included owing to uncertainties associated with difficulty in crystal growth, interstitial oxygen doping, and coexisting spin density wave structure (highest $T_C \approx 30$ K occurs for La-based epitaxial films; we estimate $T_{C0} \approx 32$ K based on known Nd-based structure) [116]. The p-type 1111 iron pnictides, e.g. $(Pr_{1-x}Sr_x)FeAsO$ ($T_C \approx 15$ K) [117], were not included because they appear to be non-optimal, likely due to the comparatively large mismatch in ionic radii between the Lanthanides (e.g. the ionic radius of $Pr^{+3}$ is 0.99 Å) and $Sr^{+2}$ (ionic radius = 1.18 Å), which would induce scattering (in the absence of such scattering we estimate $T_{C0} \approx 40$ K). The 111 and 11 Fe pnictides such as $Li_xFeAs$ and $FeSe_{1-1/8}$ (or $Fe_{1+1/8}Se$) were also not included in the analysis herein since the required structural (doping) asymmetry is apparently accomplished via vacancies, interstitials or inclusions (see e.g. [118]), which are difficult to quantify. The layered transuranium compounds such as $PuCoGa_5$ ($T_{C0}$ = 18.5 K, which is readily depressed by disorder from self-radiation damage [119]) were excluded because the absence of anions and mixed valences make direct comparison to $YBa_2Cu_3O_{6.92}$ rather speculative.[30] Finally, materials based on molecular forms of carbon (e.g. $A_3C_{60}$) may show resemblances to high-$T_C$ superconductors and will therefore be considered in future study.

To prove the validity of our model, it was first necessary to accurately determine the $\ell$ and $\zeta$ values of various high-$T_C$ compounds. Determining $\zeta$ is straightforward, and owing to features (structure type and valence) shared with $YBa_2Cu_3O_{6.92}$, $\ell$ for many of the high-$T_C$ compounds is also well defined. For the remainder, we use a set of simple rules based on shared doping, assuming that carrier density scales with the ionic valence. The result given in figure 2 shows that the transition temperatures for all 31 of the optimal high-$T_C$ compounds from several families obey the universal relation defined in equation (2.6), pinning the energy of the pairing interaction to the 2D charge density per type I layer and the spatial separation between nearest-neighbor cations and anions. It is important to note that knowing the loci of the superconducting condensate and the mediating carriers is not a requirement in producing figure 2. The theoretical model as formulated simply requires two types of physically separated layer structures, hosting two types of carriers or bands (presumably of opposite sign, although same-sign carriers fall within the purview of the theory).

[30] It is possible to speculate on how our rules may be applied to $PuCoGa_5$ by mapping this system (with presumed valences of $Pu^{+3}$, $Co^{+2}$ and $Ga^{+3}$ and associating $Pu^{+3}Ga^{+3}$ with the $Cu^{+2}$ ions of the $CuO_2$ layers) onto an analogous cuprate structure: $(Pu^{+3}Ga^{+3})$ has the carrier equivalency of $Cu^{+2}$, $(Ga^{+3})_2$ is equivalent to $Ba^{+2}$ and $Co^{+2}$ is equivalent to $Cu^{+2}$, all reflecting an equivalent cuprate cation structure of the $Ba_2Cu_2$, with $\eta = 1$. Assuming that the missing oxygens contain half of the charge (rule 2.b), their absence in $PuCoGa_5$ contributes a factor of $(1/2)(10_{oxy}/14_{oxy})$, and since the lost charge associated with the virtual anions, $(1/2)(10/14)\ \sigma_0$, must be shared among 4 cation layers (rule 1.a) one obtains $\sigma = (1/2)(10/14)(1/4)\ \sigma_0 = 0.0893\ \sigma_0 = 0.0204$. Given $A = 18.1084$ Å$^2$ and $\zeta = 2.0966$ Å (distance between $Pu_z$ and $Ga_z$), equation (2.6) yields $T_{C0}$ = 19.9 K, which is within 1.4 K of the measured value.

With this understanding, and following a strong-coupling formalism, we provide a theoretical confirmation of our experimental findings based on Coulomb interactions between spatially separated layers of charge carriers. In section 4.1 we discuss the nature of the paring mediation that is gleaned from a variety of experimental observations. In section 4.2 we explore the role (if any) of the effective mass in determining $T_{C0}$ and the possible existence of a universal superconducting reduced effective mass (as deduced from penetration depth measurements) for the high-$T_C$ compounds. In section 4.3 we put forth a Compton scattering mechanism as a possible way of understanding the apparent insensitivity of $T_{C0}$ to band mass, and in section 4.4 we discuss our pairing model in the limit of thin films where the thickness approaches the periodicity *d*.

### *4.1. Nature of the pairing mediation*

Application of the universal result for $T_{C0}$ in equation (2.6) across a diversity of compounds clearly validates identifying the type I and type II structures (figure 1) with the superconductive pairs and charges most intimately involved in the spatially indirect attractive interaction. Let us explore the implications of drawing a distinction between charges in the superconducting condensate and charges responsible for mediating the pairing interaction by examining experiments for the well-studied exemplary optimum compound, $YBa_2Cu_3O_{7-\delta}$ (noting that proximity and scattering effects may impart dual character to charges and carriers in both structural types; in some layered models the same charges carry and mediate the superconductivity [120]). Non-superconducting and marginally non-metallic components have been observed in various experiments, which indicate involvement of normal-like carriers in the electronic excitations that mediate the superconductivity of the holes. Notable observations are the anomalous residual ac conductivity in the limit of zero frequency and the incoherent transport along the *c* axis [121], which illustrate that strongly scattered carriers with three-dimensional mobility are present in the superconducting state. Appearance of normal-like components coexisting with the superconducting condensate is also consistent with measurements of excess thermal conductivity and specific heat, as noted earlier, where doping dependence in the latter case implies three-dimensionality in the relevant density of states. Transport in the normal state along the *c*-axis (resistivity and Hall coefficient) reveals that these non-metallic carriers behave like negatively signed electrons [122]. The picture synthesized from this and other evidence is one of superconductive pairing of holes mediated by an interaction involving electrons of distinctly different character [19]. Strong scattering and three-dimensionality of the electrons (but not the holes) are important for breaking electron-hole symmetry (thereby diminishing the possibility of forming electron-hole bosons such as proposed in [123]). One may therefore presume that sustained superconductivity in the electrons is suppressed (i.e. weakly contributing to macroscopic supercurrent flow and vorticity); however, owing to their proximity to superconducting holes, paired electrons may exhibit transient superconductive coherence. Certain characteristics of the iron pnictides bear striking similarity to this view of the cuprates. Of relevance here is the coexistence of two electronic subsystems determined from infrared reflectivity investigations, and the finding of a temperature independent incoherent background component in the optical conductivity [124].

From the charge allocation rules applied in determining equation (2.6) one might infer that for $YBa_2Cu_3O_{7-\delta}$ the holes are in the type I layers and the electrons are in the type II layers; however, the theoretical result presented herein is completely general, with no required assignment as to the location of the superconducting condensate to this point. Thus, one must appeal to independent considerations to determine the proper assignments. Indeed, for the Cu-doped ruthenate $Sr_2Y(Ru_{0.9}Cu_{0.1})O_6$, it was shown that the superconducting hole condensate must reside in the type I SrO layers; fluctuating Ru moments freeze at a Néel temperature of $T_N = 23$ K in an antiferromagnetic structure, producing a local field on the order 3 kG at the muon site in the type II $(Y,Ru_{0.9}Cu_{0.1})O_2$ layers, and a null magnetic field in the type I layers (second muon site) [125] such that $T_N = T_C = 23$ K. Thus, the mediating electrons would have to reside in the $(Y,Ru_{0.9}Cu_{0.1})O_2$ structures, where their interactions with the static local field would presumably have little or no effect on the superconducting state. In the case of $YBa_2Cu_3O_{7-\delta}$ it has been proposed that the locus of the holes is in the type I (BaO-CuO-BaO) layers [126]. This is based on bond-valence sum considerations of the type I and II structure properties [127-130], observed destruction of superconductivity for rare-earths substituting for Ba in $YBa_2Cu_3O_{7-\delta}$ [5,131], the transfer of electrons away from the type I structures in underdoped $YBa_2Cu_3O_{7-\delta}$ [51], differing manifestations of pairing state symmetries [132,133], the non-requirement of $CuO_2$ planes for superconductivity [2,134],[31] and conversion to non-superconducting n-type with quadravalent substitution for $Y^{+3}$ [127].

Although cuprates and ruthenates share structural similarities, for the p-type cuprates it has been conventional to consider the type II layers to be the locus of the superconducting holes, which would then involve the type I layers in mediating the pairing interaction. This particular assignment takes into account that charges in type I layers may not be fully itinerant in all cuprates. Under this assumption the mediating charges must be associated with the $Sr^{+2}$/ $Ba^{+2}$ or equivalent cation sites. In our study of the entire cuprate family we found that that $T_{C0} \propto \zeta^{-1}$ is obeyed only by measuring $\zeta$ to the site of the positive ion; the relationship breaks down if type I O-ion sites are used, as it does for any apical-oxygen model [44], particularly since the n- and p-type "infinite layer" and n-type T′-structure compounds, among others, do not contain the apical type I oxygens associated with the outer cations. Moreover, one may not automatically assume that the roles of the types I and II layers are reversed for the n-type cuprates, given that differences in charge allocation associated with $CuO_2$ planes has been observed [116]. The F-doped 1111 iron-pnictide superconductors also can be either n- or p-type [13,14]. These systematic similarities suggest that high-$T_C$ superconductors, particularly those with mixed valences, are generally ambipolar with respect to the character of their types I and II layers.

### *4.2. $T_{C0}$ dependence on effective mass*

The most surprising feature of the data is that the optimal transition temperature $T_{C0}$, which is accurately determined by equation (2.6), appears to be independent of an effective mass, i.e. it does not contain effective masses of either the pairing or the mediating carriers. This absence could imply independence of the superconducting mechanism on this aspect of the band structure (a rather unique concept). Perhaps a more tractable possibility is weak variation of effective masses among high-$T_C$ compounds in their superconducting states. One such effective mass, denoted here as $m_\lambda^*$, is determined by the in-plane London magnetic penetration depth $\lambda_\parallel$ and the carrier concentration through the relation,

$$\lambda_\parallel^2 = m_\lambda^* d c^2/(4\pi e^2 n_{2D}), \qquad (4.1)$$

where the carrier density $n_{2D}$ is the same for both type I and type II charge reservoirs. From measured (approximate) values of $\lambda_\parallel$ it is possible to utilize this relationship in combination with equation (2.3) to determine the behaviour of the effective mass using the equation,

$$m_\lambda^*/m_0 = 4\pi\,(e^2/m_0c^2)\, n_{2D}\lambda_\parallel^2/d \approx (3.541 \times 10^{-4}\ \text{Å})\, \lambda_\parallel^2 \sigma\nu/dA\ , \qquad (4.2)$$

where $e^2/m_0c^2 \equiv r_e \approx 2.818 \times 10^{-5}$ Å is the classical electron radius; values for $d$, $\sigma$, $\nu$, and $A$ are given in table 1 and $\lambda_\parallel$ is determined from experiment. For $YBa_2Cu_3O_{6.95}$ ($\lambda_\parallel$ = 1276±15 Å [33,132]), $Bi_2Sr_2CaCu_2O_{8+\delta}$ ($\lambda_\parallel$ = 2050±150 Å [135]), $La_{1.837}Sr_{0.163}CuO_4$ ($\lambda_\parallel$ = 2185±100 Å [2]) and κ–[BEDT-TTF]$_2$Cu[NCS]$_2$ ($\lambda_\parallel$ = 7680±700 Å [136]; $\sigma$ and $\nu$ are the same as for κ–[BEDT-TTF]$_2$Cu[N(CN)$_2$]Br; $d$ and $A$ are taken from [2]), one obtains $m_\lambda^*/m_0$ = 1.51±0.04, 1.50±0.22, 1.47±0.13 and 1.41±0.26 with $T_{C0}$s of 93.7, 89, 38 and ~10 K, respectively (corresponding to an average effective mass ratio of 1.50±0.21). For n-type $Ba(Fe_{1.84}Co_{0.16})As_2$, $\lambda_\parallel$ = 3600±500 Å [137], and $m_\lambda^*/m_0$ = 1.8±0.5, in agreement with the p-type compounds. These results are consistent with the hypothesis of rather small variability in carrier mass (in the superconducting state) and with values close to the bare electron mass $m_0$.

---

[31] Early papers advocating the BaO layers as the locus of the superconducting hole condensate in the cuprates did not include possible superconductivity of the $CuO_2$ planes.

Properties that scale with density of states, such as specific heat and tunnelling, are more likely to reveal characteristics of the mediating charges. For example, analysis of the jump in the specific heat at $T_C$ was used to determine a specific heat effective mass $m_\gamma$*, which for $YBa_2Cu_3O_{7-\delta}$ is (12.0 ± 2.4) $m_0$ and for $Bi_2Sr_2CaCu_2O_{8+\delta}$ is (7.8 ± 2.4) $m_0$ [2]. Although these values are considerably greater than the transport $m_\lambda$* given above, they are consistent with $m_e^* >> m_h^*$ (effective masses of electrons and holes, respectively). We note here that $m_\gamma^* \sim m_e^* + m_h^*$, whereas $m_\lambda^* \sim (1/m_h^* + \tau/m_e^*)^{-1}$, where $\tau < 1$ is introduced as a scattering correction. In this interpretation, one has $m_\gamma^* > m_\lambda^*$, which is in agreement with observation. Electrons of larger mass possess the higher density of states and have the greater tunnelling probability (as compared to the lighter holes); therefore the electron component could dominate the flow of current through weak link structures.

Our result of equation (2.6) also provides new insight on why any analysis using only penetration depth components $\lambda_\parallel$ fails to yield a predictive formula for $T_{C0}$, as was demonstrated for several cuprate families in [138] (a study of doping dependence that includes optimum compounds) and various Fe pnictides in [139]. The expression in equation (4.1) involves the three-dimensional carrier density $n_{2D}/d$, which is a reminder that a measurement of $\lambda_\parallel$ does not fully capture the distinctive 2D character of high-$T_C$ superconductivity. Even if one were to ignore realities that $T_{C0}$ depends on $\zeta$ and that $\lambda_\parallel$ depends on m*, the 2D superfluid density obtained from $\lambda_\parallel$ does not reveal the correct interaction density $\ell^{-2}$ that applies to compounds at optimum doping.

### *4.3. Pairing via Compton scattering of virtual photons*

The apparent absence of any effective mass dependence in equation (2.6) might also be understood by considering the energy scale of the pairing interaction. Recognizing that the fundamental length $\Lambda$ introduced in equation (2.8) is (to within 3%) equal to twice the reduced electron Compton wavelength $\bar{\lambda}_e = \hbar/m_0c \approx 0.00386$ Å [140], equation (2.8) may also be expressed as,

$$k_B T_{C0} \approx 2\frac{\bar{\lambda}_e}{\ell}\frac{e^2}{\zeta} \quad . \tag{4.3}$$

Presented in this form, our theoretical result illustrates that $T_{C0}$ scales with the Coulomb energy $e^2/\zeta$, reduced by a factor containing the ratio of Compton wavelength to superconducting interaction length. This form contains only the electron rest mass, emphasizing independence of band masses. In addition to verifying that the Coulomb interaction is responsible for high-$T_C$ superconductivity, the appearance of $\bar{\lambda}_e$ in this theoretical expression for $T_{C0}$ suggests that virtual photon interactions are also playing a role ($e\mathbf{p}\cdot\mathbf{A}/mc$ term in Hamiltonian, Coulomb gauge). Accordingly, we show in figure 5 a scattering diagram that includes both the collective electronic excitations and virtual photon exchange in the superconductive pairing interaction. Pairing particles $h_k$ and $h_{-k}$ indirectly interact by creating and annihilating virtual photons of frequency and wave vector $\nu_q$ and $\nu_{-q}$, respectively, which in turn scatter electronic excitation $e_\kappa$ into virtual state $e_{\kappa+q}$ before returning to state $e_\kappa$. Scattering by virtual photons are shown explicitly because of the scaling of $T_{C0}$ with $\bar{\lambda}_e$ and the derivation of equation (3.6) that presumes high energy scattering, $\hbar\nu_q \sim E_F$.

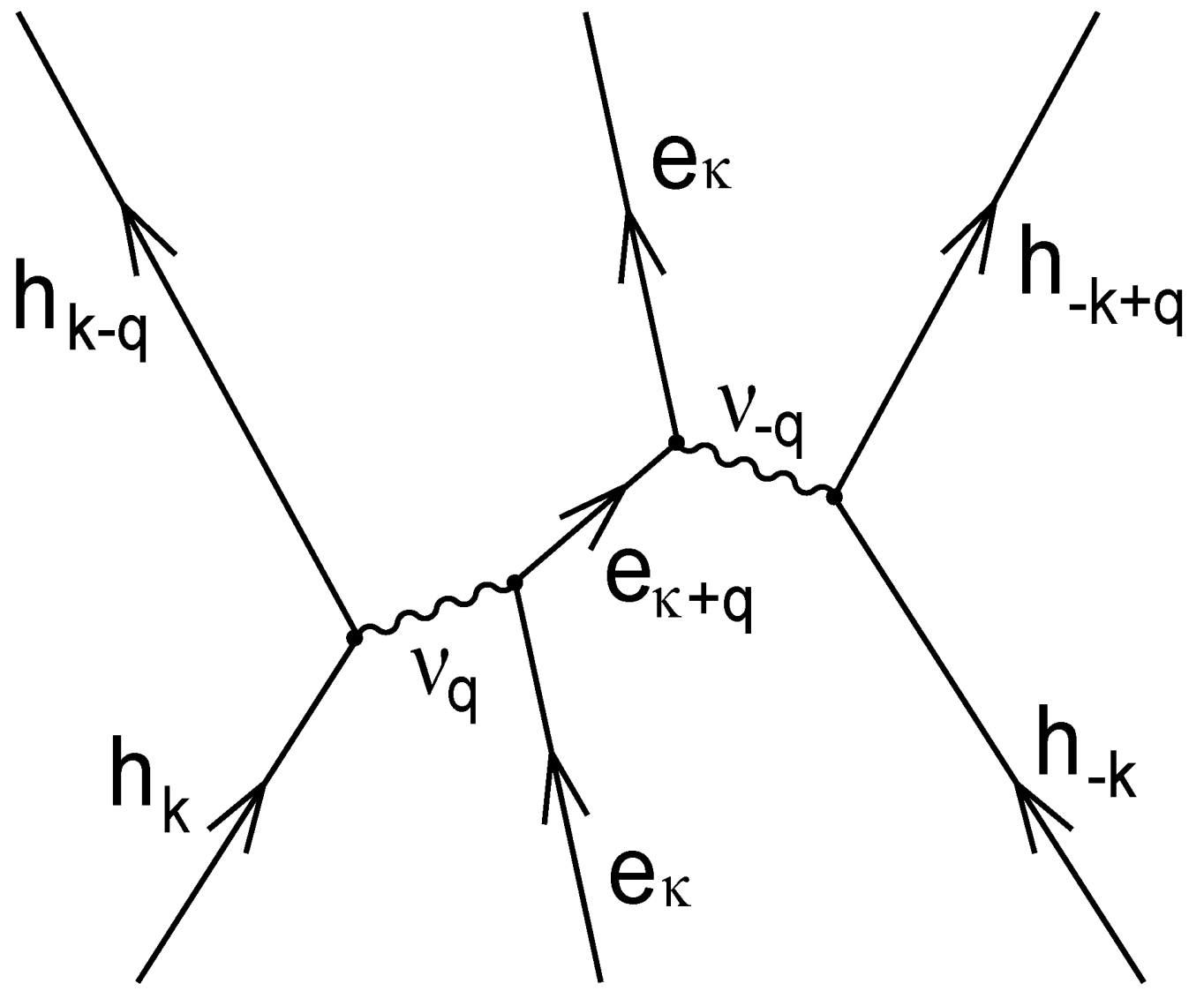

**Figure 5**. Interaction diagram of high-$T_C$ pairing mechanism, where h, e, and ν denote holes, electronic excitations, and photons, respectively; k, κ, and q denote momenta.

Although superconducting transition data for each compound were carefully evaluated for this work, there are small uncertainties in $T_{C0}$ from the width of the superconducting transition and possible deviation of some samples from absolute optimum. Samples with $T_C < T_{C0}$ would systematically bias the fitted constant β and the result for Λ from equation (2.7) toward smaller values. It is of interest for our future work to extend the theory, which presently determines $T_{C0}$ only for the (unperturbed) optimum compounds, by treating the variation of $T_C$ with uniaxial pressure and the non-linear variation of $T_C < T_{C0}$ with doping.

### *4.4. The elemental high-$T_C$ superconducting structure*

High-$T_C$ superconductor structure, stylized in figure 1, comprises multiple layers that repeat a unit building block of thickness equal to the periodicity distance *d*. This elemental unit contains all the components, a complete set of type I and type II structures, necessary to induce superconductivity, with the pairing mediated through interlayer Coulomb interactions. It is, therefore, of interest to consider a thin crystal of thickness *d* obtained from bulk crystal structure in the form of a section extended in the basal plane and truncated in the transverse direction. Such a structure is predicted to be superconducting, assuming reasonable replication of bulk-like boundary conditions at the terminating ionic monolayers as well as maintaining the components of the type I and type II charge reservoirs required for charge transport and equilibration. A recent study of several cuprates has shown that the minimum thickness of such a superconducting ultra-thin crystal corresponds to one formula unit (per basal-plane area) [141].

In the extreme abstract limit of a conducting sheet in isolation from adjacent charge layers, the mediating (interlayer) Coulomb pairing interaction (i.e. the mechanism for high-$T_C$ superconductivity advocated herein) is removed. A well-studied example drawn from the cuprates is the $CuO_2$ monolayer (e.g. the complete type II structure that is contained in cuprates such as $La_{1.837}Sr_{0.163}CuO_{4-\delta}$ and $HgBa_2CuO_{4.15}$). From theoretical considerations of spin plaquette structures in the underdoped pseudogap regime and formation of loop-ordered states in $CuO_2$ monolayers, Aji et al. have found that coupling of quantum critical fluctuations to the conduction-band carriers leads to superconductive pairing with *d*-wave symmetry [142]. Their results are consistent with the occurrence of maximum $T_{C0}$ at optimal doping (near the quantum critical point), but can provide only order-of-magnitude values for $T_{C0}$ for these simpler cuprates. In this theoretical picture, the type I doping layers serve a passive role, providing the free carriers residing in the $CuO_2$ structure.

## 5. Conclusion

The transition temperatures of high-$T_C$ superconductors were shown to depend systematically on a layered structure comprising two distinct charge reservoirs, denoted as type I and type II, that contain charge densities $\sigma_I \nu/A$ and $\sigma_{II}\eta/A$, respectively, and the distance $\zeta$ between nearest-neighbor type I and type II interacting layers. Accepting that charge transfer along the hard axis allows equal charge densities in the two reservoirs at equilibrium, we obtain a definition for areal charge density $n_{2D} = \sigma\nu/A$ and the superconducting interaction density $\ell^{-2} = \sigma\eta/A$ ($\sigma \equiv \sigma_I$). The charge fractions $\sigma$ are calculated by analyzing shared doping and ionic valences. Using $YBa_2Cu_3O_{6.92}$ as a reference, and introducing a set of charge allocation rules, we were able to calculate $\zeta$ and $\ell$ for 31 optimal compounds (recognizing that only the optimal high-$T_C$ compounds possess intrinsic bulk superconducting properties) and show that the transition temperatures follow the universal function: $T_{C0} = k_B^{-1}(\beta/\ell\zeta)$. The remarkably good agreement between theory and data (ocular proof in figure 2) confirms the validity of equation (2.6) and the Coulombic nature of the high-$T_C$ pairing mechanism. Moreover, expression of the result $k_B T_{C0} = e^2\Lambda/\ell\zeta$ [equation (2.8)] was shown to follow from the root-mean-square average of indirect (interlayer) Coulomb forces. Note that the theory was developed absent any designation as to the locus of the superconducting condensate. However, it does appear that for $Ba_2YRu_{0.9}Cu_{0.1}O_6$, certain pnictides and the BEDT-TTF based organics, the dominant superconducting condensate resides in the type I layers.

A particularly intriguing finding is the apparent independence of $T_{C0}$ on any band mass or various Fermi surface topologies. From this we conclude that $T_{C0}$ depends (to zeroth order) on neither the dispersion of the energy bands nor their proximity to the Fermi level (i.e. on whether charges are to be regarded as itinerant holes or electrons, or as valence-state charges). We show in equation (4.3) that $T_{C0}$ is given by the Coulomb potential $e^2/\zeta$ modified by a factor containing the ratio of the reduced Compton (electron) wavelength $\lambda\!\!\!^{-}_e$ to superconducting interaction length $\ell$, thereby providing a clue to the high-$T_C$ mechanism. Since $\lambda\!\!\!^{-}_e$ contains only the electron rest mass, this expression of our finding emphasizes the independence of $T_{C0}$ on band masses. In addition to verifying that the indirect Coulomb interaction is responsible for high-$T_C$ superconductivity, the appearance of $\lambda\!\!\!^{-}_e$ in the theoretical expression for $T_{C0}$ suggests that virtual photon interactions play a role.

Our results also provide a straightforward explanation for the absence of a scaling relationship between $T_{C0}$ and superfluid density (e.g. as derived from the London penetration depth). $T_{C0}$ involves the superconducting interaction areal density $\ell^{-2}$ and the interaction length $\zeta$, whereas the three-dimensional superfluid density involves $n_{2D}$ and the periodicity $d$. Since in general one finds from the data in table 1 that $\eta/\nu$ and $\zeta/d$ are not universal constants, there is no universal relationship between $T_{C0}$ and superfluid density. In the universal relationship discovered in this work, $T_{C0}$ is determined by the elemental high-$T_C$ superconducting structure comprising the types I and II charge reservoirs.

## Acknowledgements

We are grateful for the support of Physikon Research Corporation (Project No. PL-206), and the New Jersey Institute of Technology. We note with interest that 2011 is the 100th anniversary of the discovery of superconductivity [143,144] and the 25th anniversary of the discovery of high-$T_C$ superconductivity [1]. Publications on this work have appeared [145-147] with corrigendum [146] incorporated herein.